\documentclass[final,5p,times,twocolumn]{elsarticle}
\usepackage{amsmath,amssymb,amsfonts}%
\usepackage{amsthm}%
\usepackage{graphicx}%
\usepackage{multirow}%
\usepackage{mathrsfs}%
\usepackage[title]{appendix}%
\usepackage{xcolor}%s
\usepackage{textcomp}%
\usepackage{manyfoot}%
\usepackage{longtable}
\usepackage{tabularx}
\usepackage{graphicx} % for \resizebox
\usepackage{booktabs}%
\usepackage{rotating}
\usepackage{algorithm}%
\usepackage{algpseudocode}%
\usepackage{listings}%
\usepackage{pdflscape} % or \usepackage{lscape}
\usepackage{pifont} % This is the crucial package for \xmark and \cmark
\usepackage{longtable}
\usepackage{array}
\usepackage{color}
\usepackage{caption}
\usepackage{url}
\usepackage{natbib}
\newcommand{\cmark}{\ding{51}}  % Tick
\newcommand{\xmark}{\ding{55}}  % Cross
\usepackage{tikz}
\usepackage{tabularx}
\usepackage{tabularray}
\usepackage{pdflscape}
\usepackage{hyperref}
\usetikzlibrary{shapes, arrows.meta, positioning}

\tikzset{
	stage/.style={rectangle, draw=blue!60, fill=blue!20, thick, minimum width=3.2cm, minimum height=1.2cm},
	subtask/.style={ellipse, draw=gray!60, fill=gray!20, thick, minimum width=2.8cm, minimum height=0.8cm},
	arrow/.style={-{Stealth}, thick}
}

\begin{document}
	
	\begin{frontmatter}
		
		\title{CloudyGUI: A Novel Python-based Framework for Auto-Scaling and Cloud Workload Analysis}
		
		\author{Jyoti Bawa*} %% Author name
		
		\affiliation{organization={Guru Nanak Dev University},%Department and Organization
			addressline={Department of Computer Science},
			city={Amritsar},
			postcode={143005},
			state={Punjab},
			country={India}}
		
		\author{Mohit Kaushik}
		\affiliation{organization={Guru Nanak Dev University},
			addressline={Department of Computer Science},
			city={Amritsar},
			postcode={143005},
			state={Punjab},
			country={India}}
		
		\author{Kuljit Kaur Chahal}
		\affiliation{organization={Guru Nanak Dev University},
			addressline={Department of Computer Science},
			city={Amritsar},
			postcode={143005},
			state={Punjab},
			country={India}}
		
		\author{Kamaljit Kaur}
		\affiliation{organization={Guru Nanak Dev University},
			addressline={Department of Engineering and Technology},
			city={Amritsar},
			postcode={143005},
			state={Punjab},
			country={India}}
		
\begin{abstract}
		
		{\textbf{Purpose:} Cloud computing environments are highly dynamic, creating major challenges for resource management. Accurate workload prediction is therefore essential for effective auto-scaling. To address this, we present CloudyGUI, a Python simulation framework with an easy-to-use GUI that allows researchers to test and validate resource management strategies. \textbf{Methods:} This framework employs a three-stage pipeline: workload generation, prediction (utilizing XGBoost and LSTM), and an auto-scaling system based on the MAPE loop. Validation includes internal, intermediate, and external methods to ensure system reliability. \textbf{Results:} CloudyGUI's generated workloads closely match real-world datasets. A two-sample K-S test confirms this alignment, showing strong p-values of 0.19 for CPU and 0.14 for memory. When compared to a command-line tool, the GUI adds only a minimal overhead of 1.4×-4.67×. Furthermore, expert review validates the tool's realism and practical usefulness. \textbf{Conclusion:} CloudyGUI fills a critical gap by providing an accessible and efficient platform for simulating auto-scaling in cloud applications, helping researchers develop advanced cloud management solutions.}

\end{abstract}
		
		\iffalse
		\begin{graphicalabstract}
			%\includegraphics{grabs}
		\end{graphicalabstract}
		
		\begin{highlights}
			\item Research highlight 1
			\item Research highlight 2
		\end{highlights}
		\fi
		
		
		\end{frontmatter}
		
		% --- arXiv Acceptance Note ---
		\let\thefootnote\relax\footnotetext{\textbf{Published in Simulation Modelling Practice and Theory. Cite the published version: \url{https://doi.org/10.1016/j.simpat.2026.103308}}}
		
		\setlength{\abovecaptionskip}{10pt}
		\setlength{\belowcaptionskip}{10pt}
		\setlength{\textfloatsep}{10pt}
		
		\section{Introduction}\label{sec1}
		Cloud environments are highly complex due to virtualization, multi-tenancy, and auto-scaling~\cite{mansouri2020cloud,cai2017elasticsim,rout2023dynamic,sabry2011cloud}. This makes it difficult for researchers to test their policies in a real cloud environment~\cite{mansouri2020cloud}. Researchers often use simulation tools to analyze cloud performance and resource behavior without needing actual cloud systems~\cite{cai2017elasticsim,asir2018analysis}. However, traditional simulators often fail to accurately replicate the dynamic behavior of the cloud environment~\cite{mansouri2020cloud,ngharamike2018cloud,umar2023simulation,pandey2014comparative,siavashi2024cloudy,kapil2024cloud}. To overcome this, we need a specialized simulation tool that can accurately capture these complexities. Such tools are essential for evaluating strategies for resource provisioning, load balancing, and energy efficiency~\cite{sajitha2018analysis,kathiravelu2014adaptive}. Thus, researchers can run repeated experiments without the expense of real-world cloud deployment~\cite{maarouf2015comparative,mansouri2020cloud,Goga2014}.

		Existing tools are mostly built in Java and C++, with CloudSim serving as a key resource~\cite{shahid2023systematic,kumar2019issues}. However, these Java-based simulators often face difficulties related to adaptability and flexibility, which are necessary to accurately simulate the complexities of modern cloud environments~\cite{fakhfakh2017simulation,siavashi2024cloudy}. Nowadays, a major challenge involves implementing and evaluating various auto-scaling techniques~\cite{fakhfakh2017simulation,sakellari2013survey,bashar2014modeling}. 
		
		To address these limitations, more flexible and modern simulation tools are needed~\cite{maarouf2015comparative,shahid2023systematic,kumar2019issues}. Tools like Cloudy~\cite{siavashi2024cloudy} recently started a trend toward Python-based solutions. Python is preferred for its simplicity, readability, and easy integration with various libraries for GPU computing. However, This transition has been slow, and several challenges remain. While Python offers features like simplicity and integration with AI/ML libraries, existing efforts often resulted in minimal innovation, such as projects that primarily translated older Java frameworks (e.g., geoCloudSim\footnote{\url{https://github.com/AUT-Cloud-Lab/geoCloudSim}}) or became inactive for extended periods (e.g., pyCloudSim\footnote{\url{https://github.com/vonpupp/pyCloudSim}}). These issues underscore the existing gap for a comprehensive, natively designed, and functionally advanced Python simulation framework capable of handling auto-scaling and providing enhanced usability.

		To address this gap, this paper introduces a novel Python-based cloud simulation framework. The key contributions of this work are as follows:
		\begin{itemize}
			\item \textbf{A Python-Based Simulation Framework:} We present a Python-based cloud simulator that directly addresses the limited evolution and provides ease of use.
			\item \textbf{Graphical User Interface for Enhanced Usability:}  The framework includes a user-friendly GUI, significantly lowering the barrier to entry for setting up, running, and analyzing complex cloud simulations.
			\item \textbf{Predictive Auto-Scaling:}  We implement a predictive threshold-based auto-scaling mechanism, a critical feature for modern cloud environments that is underdeveloped in existing Python-based simulators.
		\end{itemize}
		
	    While simulators with graphical interfaces and auto-scaling exist~\cite{nunez2012icancloud,li2012dartcsim,aslanpour2021autoscalesim,papadopoulos2016peas,kim2015pics,cai2017elasticsim}, they remain largely Java-based. In contrast, modern resource management research is driven by Python-based ML frameworks, creating a disconnect between the two environments.
	    This architectural mismatch forces researchers to bridge or rewrite ML models for simulation, introducing latency and complexity~\cite{siavashi2024cloudy}. CloudyGUI's key innovation is not merely its Python implementation, but its ability to execute native ML models (e.g., XGBoost, PyTorch) directly within the auto-scaling loop. 
	    
		The rest of the paper is organized as follows: Section~\ref{sec2} reviews the existing literature to identify the gaps. Section~\ref{sec3} details the methodology employed in our proposed simulator. In Section~\ref{sec4}, we present our results and analysis, demonstrating the simulation data using visualizations. Section~\ref{sec5} interprets these results by discussing their implications for existing practices. Section~\ref{sec6} concludes this study by summarizing the contributions. Finally, Section~\ref{sec7} outlines potential future directions and addresses the limitations of our study.
		
		\section{Literature Review}\label{sec2}
		Cloud computing offers features such as elasticity, cost-efficiency, and scalability~\cite{segun2024assessing}. However, this complex environment presents challenges in dynamic resource management, energy efficiency, and load balancing~\cite{shri2024optimizing,Reddy2023Efficient}. Because of high costs, risks, and time consumption, it is impractical to perform experimentation in live cloud systems~\cite{alshathri2020comparative}. Therefore, robust simulation tools are essential for understanding the complex cloud environment~\cite{siavashi2024cloudy,sanjalawe2023cloud}. These tools let users perform experiments cost-effectively~\cite{fakhfakh2017simulation}. Auto-Scaling is a strategy for dynamically managing cloud resources, which can be effectively studied using simulation. Auto-scaling helps to manage resources and adjust them according to user demand. It ensures high resource availability, maintains cloud performance, controls costs, and prevents both over-provisioning and under-provisioning in the system~\cite{vemasani2024AchievingAT}. While traditional approaches often rely on static thresholds, recent advances on control methods have shifted towards intelligent automation. Techniques utilizing AI automation~\cite{karamthulla2023optimizing}, dynamic load balancing models~\cite{rout2023dynamic}, and deep learning-based proactive scaling~\cite{taha2024proactive} are increasingly adopted to handle workload volatility. These advanced control strategies require robust simulation environments to validate their stability and performance before deployment.
		
		The evaluation of such complex cloud systems often relies on simulation tools. CloudSim~\cite{silva2017cloudsim} is a widely used tool for cloud simulations that supports the modeling of large-scale infrastructures. However, it lacks native support for auto-scaling mechanisms and a graphical user interface. Many extensions like CloudSim Plus~\cite{aslanpour2021autoscalesim}, ContainerCloudSim~\cite{piraghaj2017containercloudsim}, and GPUCloudSim~\cite{siavashi2019gpucloudsim} enhance its capabilities for modern cloud features but do not fundamentally alter its core limitations regarding auto-scaling. Other simulators like DCSim~\cite{tighe2012dcsim} focus on Virtual Machine (VM) management and dynamic resource allocation, modeling multi-tier applications and supporting features like VM migration and overcommitted resource provisioning. Meanwhile, MDCSim~\cite{lim2009mdcsim} targets multi-tier data centers, featuring a pluggable three-layer architecture to model communication protocols, kernel-level scheduling, and user-tier interactions, with validation against real prototypes.
		
		Several simulators have incorporated auto-scaling, primarily for web applications, though each has unique focuses and limitations. AutoScaleSim~\cite{aslanpour2021autoscalesim}, an extension of CloudSim, offers a comprehensive toolkit for evaluating auto-scaling in web applications, supporting the full MAPE-K loop, customizable simulations, and various performance metrics, even allowing real web traffic testing and OpenStack validation. However, it largely focuses on generic web server workloads rather than the bursty patterns of interactive GUI applications. PEAS~\cite{papadopoulos2016peas} evaluates auto-scaling using scenario theory with probabilistic guarantees and real-platform validation, precisely measuring low-level metrics but overlooking high-level aspects like failed requests, SLA compliance, and cost for generality. PICS~\cite{kim2015pics} provides an auto-scaling framework for public clouds, considering cost and SLA from an end-user perspective, but it primarily simulates workflow rather than transactional workloads and is restricted to short-term simulations (minutes), insufficient for long-term interactive user experiences. Addressing static scheduling limitations, ElasticSim~\cite{cai2017elasticsim}, a CloudSim-based toolkit, simulates resource auto-scaling with variable task execution times, using probability distributions and interval-based VM pricing to show that stochastic execution times can significantly increase costs and lead to deadline violations, emphasizing the need for dynamic, uncertainty-aware scheduling.
		
		Beyond the aforementioned tools, several specialized simulation toolkits extend cloud modeling capabilities for diverse scenarios, though often without specific focus on the unique demands of interactive GUI auto-scaling. SPECI~\cite{sriram2009speci} focuses on the scalable design, performance, and failure/recovery of cloud data centers, allowing users to explore these aspects by defining data center size and middleware policies. For scientific applications, GroudSim~\cite{ostermann2010groudsim}, a Java-based toolkit enabling simulation on Grid and cloud infrastructures, providing performance statistics and modeling computational and network aspects, including job and file transfers. MDCSim~\cite{lim2009mdcsim}, a commercial and scalable toolbox, offers in-depth analysis of multi-tier data centers by modeling hardware for power consumption estimates and optimizing web application performance through various resource configurations and network topology. Furthermore, NetworkCloudSim~\cite{garg2011networkcloudsim} extends CloudSim with a scalable network and a generalized application model for cloud data centers, supporting inter-communicating elements like Message Passing Interface (MPI) and workflows, featuring a network flow model for bandwidth sharing, and allowing easy topology modification via a configuration file. While valuable for their respective domains, these tools typically lack the fine-grained control and workload modeling necessary for simulating responsive auto-scaling tailored to individual user interactions within a GUI.
		
		Furthermore, various toolkits offer advanced modeling capabilities for modern cloud aspects, yet frequently lack the specific integration of auto-scaling with interactive GUI workload characteristics. DCSim~\cite{tighe2012dcsim} provides an extensible framework for investigating dynamic resource management in IaaS, featuring multi-tier application models and VM interaction/replication. Extending CloudSim, ContainerCloudSim~\cite{piraghaj2017containercloudsim} models and evaluates containerized cloud environments, supporting CaaS, container lifecycle, and energy-aware provisioning. Similarly, ICARO~\cite{badii2016icaro} Cloud Simulator (ICLOS) focuses on long-term prediction of complex cloud workloads and business configurations by integrating a semantic Knowledge Base and Smart Cloud Engine for realistic, SLA-driven simulations. GPUCloudSim~\cite{siavashi2019gpucloudsim} is designed to evaluate GPU virtualization schemes and resource provisioning. It makes use of the First Fit Increasing (FFI) VM placement algorithm, which further improves performance and energy efficiency. Although it addresses the limitation of previous tools that are heavily focused on non-GPU resources, it fails to provide a simulator with auto-scaling capabilities built into the GUI.
		
		Several simulators offer a graphical user interface to simplify the complex cloud environment. Among these, iCanCloud~\cite{nunez2012icancloud} provides a GUI to configure virtual machines, data centers, and other resources. Other tools, such as DARTCSim~\cite{li2012dartcsim}, extend CloudSim by introducing a GUI to simplify  tasks like dynamic resource allocation and task scheduling. For Software-Defined Cloud Data Centers (SDN), CloudSimSDN~\cite{son2015cloudsimsdn} allows users to efficiently manage the network and traffic. These visual interfaces help to reduce the learning curve and also improve the user’s experience in simulation. Tools like CloudAnalyst~\cite{wickremasinghe2010cloudanalyst} and CloudReports~\cite{teixeira2014cloudreports} also offer GUIs for experimental setup but  lack robust auto-scaling.
		
		While there are numerous simulation tools, there is a lack of integration of Cloud Simulators built into Python, which comes with GUI and auto-scaling capabilities. There are many simulators, such as ICLOS~\cite{badii2016icaro} and ElasticSim~\cite{cai2017elasticsim}, which offer auto-scaling capabilities but  are built in Java. This gap limits researchers and developers who prefer Python due to its flexibility and simplicity. Existing tools and research primarily address these domains in isolation or for generic workloads, often failing to fully capture the real-time, user-centric demands and complex resource patterns of interactive cloud GUIs.

        Although tools like AutoScaleSim~\cite{aslanpour2021autoscalesim} and CloudSim~\cite{calheiros2011cloudsim} Plus support auto-scaling, they are restricted to static or pre-defined policies within a Java ecosystem, preventing the direct integration of modern Python-based AI workflows. CloudyGUI differentiates itself by treating the simulation engine as a native extension of the data science stack, rather than a standalone engineering tool.
		
		Our proposed CloudyGUI tool aims to fill this gap. CloudyGUI provides a GUI-based tool in a Python environment that handles both workload generation and auto-scaling analysis. This approach enhances accessibility and adaptability in cloud development. This integrated framework offers improved performance and user experience while optimizing resource utilization. Table~\ref{table1} summarizes the relevant literature.
		
		\section{Methodology}\label{sec3}
		There is a need for a robust and realistic simulation environment to validate various cloud management strategies. To address this, this research develops a novel simulation framework using a three-stage pipeline, which is illustrated in Figure~\ref{fig:architecture}. This pipeline shows the complete workflow, starting with workload generation and ending with the execution of predictive resource management. 
		
		The methodology begins with the workload generation process, which is designed to create datasets that accurately mimic real-world cloud environments. Next, a resource prediction framework uses machine learning to forecast future resource demands from this data. Finally, a predictive auto-scaling system integrates these forecasts into a proactive MAPE control loop for optimal resource allocation. Together, these components form a complete pipeline for rigorously testing and validating advanced auto-scaling algorithms.
		
		\subsection{Architecture Overview}
\begin{figure}
			\centering
			\includegraphics[width=\linewidth]{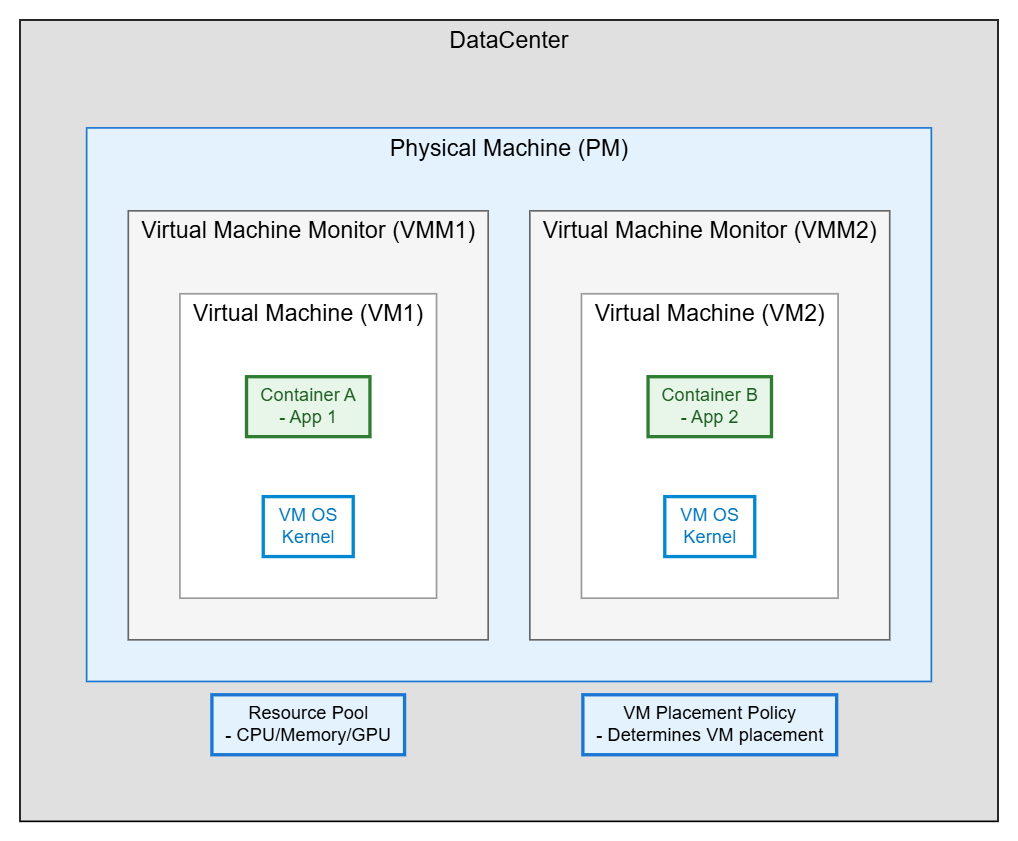}
			\caption{Conceptual Model for CloudyGUI}
			\label{fig:datacenterarchitecture}
	\end{figure}

\begin{figure*}[!h]
			\centering
			\includegraphics[width=0.8\linewidth]{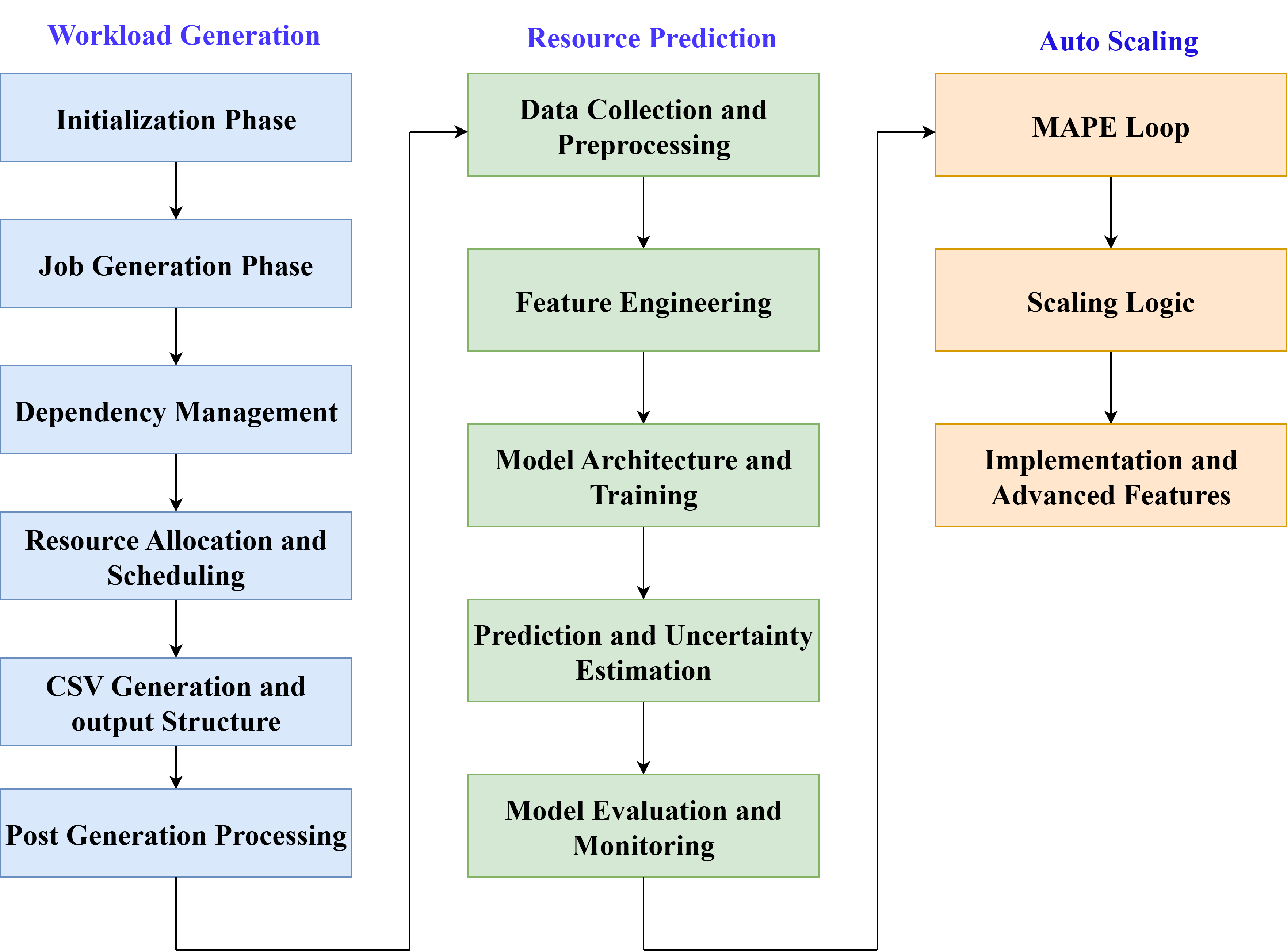}
			\caption{Architecture Diagram of CloudyGUI}
			\label{fig:architecture}
	\end{figure*}

		To effectively simulate auto-scaling behavior in cloud environments, it is required to model the underlying infrastructure with detailed modularity. Our tool adopts a hierarchical architecture that reflects the layered structure of modern cloud data centers. This allows users to control the resource allocation, virtualization and application deployment. This section presents the conceptual model used in CloudyGUI. The model shows the roles and interactions of various components within the framework. Figure~\ref{fig:datacenterarchitecture} illustrates the hierarchical structure and key components of the conceptual model.
		
		\paragraph{DataCenter (Outermost Container)}  
		The data center serves as the top-level container that holds all the elements. It includes Physical Machines (PMs), a centralized Resource Pool, and the VM Placement Policy module responsible for orchestrating virtual machine deployment.
		
		\paragraph{Physical Machine (PM)}  
		Each PM represents a physical server within the data center. It hosts multiple Virtual Machine Monitors (VMMs), which allocate physical resources such as CPU, RAM, and GPU to virtual machines. PMs are centrally managed by the DataCenter and serve as the execution substrate for virtualized workloads.

		\onecolumn
		\begin{longtable}{l c c c p{4cm} c p{4cm}}
			\caption{Literature Overview} 
			\label{table1}\\
			\toprule
			\textbf{Simulator} & \textbf{Year} & \textbf{Language} & \textbf{GUI} & \textbf{Resource Types} & \textbf{Scaling} & \textbf{Limitations} \\
			\midrule
			\endfirsthead
			\multicolumn{7}{c}{\bfseries Table \thetable{} Continued from previous page} \\
			\toprule
			\textbf{Simulator} & \textbf{Year} & \textbf{Language} & \textbf{GUI} & \textbf{Resource Types} & \textbf{Scaling} & \textbf{Limitations} \\
			\midrule
			\endhead
			% Your table content starts here
			GreenCloud~\cite{liu2009greencloud} & 2009 & C++ & \xmark & Physical Servers, VMs and traffic pattern & \xmark & Complex to configure and Limited application level abstraction \\
			
			SPECI~\cite{sriram2009speci} & 2009 & Java, C++ & \xmark & Nodes, Data, Topologies, Network Links & \cmark & Focus is limited to failure Communication and Lacks community support and general purpose extensibility \\
			
			MDCSim~\cite{lim2009mdcsim} & 2009 & Java, C++ & \xmark & CPU, Disk, Servers, Power and network modules & \cmark & No container or VM abstraction and No Fault Tolerance \\
			
			CloudSim~\cite{calheiros2011cloudsim} & 2010 & Java & \xmark & Hosts, VM, Cloudlets & \xmark & Limited support for containerization and Static workload modeling Single language support \\
			
			CloudAnalyst~\cite{wickremasinghe2010cloudanalyst} & 2010 & Java & \cmark & CPU cores. RAM. Bandwidth, VCpu, VRAM & \xmark & Outdated technology stack and No real-time simulation \\
			
			GroudSim~\cite{ostermann2010groudsim} & 2011 & Java & \xmark & Grid Sites, CPU's per Site, Job Queues, Instances, Instance Type & \cmark & Failure modeling requires manual setup and Limited community and documentation \\
			
			Network-CloudSim~\cite{garg2011networkcloudsim} & 2011 & Java & \xmark & VMs, Physical Hosts, Network Links & \cmark & Network Simulation is flow based, not packet accurate and Workload relies on user defined synthetic application models \\
			
			Workflow-Sim~\cite{chen2012workflowsim} & 2012 & Java & \xmark & Physical Hosts, VMs, tasks, overhead modules and clustering engine & \xmark & Single threaded architecture and Overhead modeling \\
			
			EMUSIM~\cite{calheiros2013emusim} & 2012 & Java & \xmark & VMs, Cloudlets, Data centers, User requests & \xmark & Workflow application emulation not natively supported, Limited cloud API integration \\
			
			CDOSim~\cite{fittkau2012cdosim} & 2012 & Java & \xmark & VMs, CPU, Memory, Bandwidth, Workloads & \xmark & Workload accuracy depends on KDM and trace quality and Limited SLA modeling \\
			
			DARTCSim~\cite{li2012dartcsim} & 2012 & Java & \cmark & Data Centers, VMs, Cloudlets & \xmark & Limited network simulation, No container support and No real-time simulation \\
			
			iCanCloud~\cite{nunez2012icancloud} & 2012 & C++ & \cmark & CPU, Memory, Disk and network & \xmark & No Built-in Power consumption modeling and Steep learning curve for customization \\
			
			FTCloudSim~\cite{zhou2013ftcloudsim} & 2013 & Java & \xmark & Physical hosts, VMs, Cloudlets, Network layers & \xmark & Cloudsim core limitations and No public cloud provider template \\
			
			DCSim~\cite{tighe2012dcsim} & 2014 & Java & \xmark & Hosts, VMs, CPU, Memory, Bandwidth, Applications & \cmark & No detailed networking layer simulation and No direct public cloud provider templates \\
			
			Cloud-Reports~\cite{teixeira2014cloudreports} & 2014 & Java & \cmark & Physical Hosts, VMs, tasks, overhead modules and clustering engine & \xmark & No container support and Energy models require manual creation \\
			
			CloudSim-SDN~\cite{son2015cloudsimsdn} & 2015 & Java & \cmark & VMs, Data centers, QOS, traffic flows & \xmark & Accuracy limited by simplifications in bandwidth sharing and latency estimation and Limitations of cloudsim \\
			
			ICARO~\cite{badii2016icaro} & 2015 & Java & \cmark & VMs, Hosts, Networks, Storage systems, SLA & \cmark & Limited Network Protocol Simulation and High Complexity for Setup \\
			
			PICS~\cite{kim2015pics} & 2015 & Java & \xmark & VMs, Storage services, Network services, Job workloads & \cmark & Manual Policy Configuration, Limited to IaaS Layer and No Network-Level Simulation \\
			
			PEAS~\cite{papadopoulos2016peas} & 2016 & Java & \xmark & VMs, Servers, CPU, Memory, Requests, Load balancers & \cmark & Assumes Known Workload Distributions and Limited to Predefined Auto-Scalers \\
			
			Container-CloudSim~\cite{piraghaj2017containercloudsim} & 2017 & Java & \cmark & VMs, Hosts, Networks, Software services, SLA & \xmark & Static resource definitions and Semantic overhead \\
			
			CloudSim Plus~\cite{silva2017cloudsim} & 2017 & Java & \xmark & Physical resources, VMs, cloudlets, power models & \xmark & Network Simulation Limitations and No container support \\
			
			ElasticSim~\cite{cai2017elasticsim} & 2017 & Java & \cmark & Workflow tasks, VMs, runtime distributions & \cmark & Limited failure modeling, and Single threaded simulation limitations No public cloud API integration \\
			
			CloudGen~\cite{koltuk2019cloudgen} & 2019 & MATLAB & \xmark & CPU cores, Memory, Usage duration & \xmark & Limited to VM-based workloads and No integration with simulators \\
			
			GPUCloud-Sim~\cite{siavashi2019gpucloudsim} & 2019 & Java & \xmark & Data centers,GPU Memory, Bandwidth, power models, VMs & \cmark & No explicit SLA modeling and Cloud native workloads \\
			
			AutoScale-Sim~\cite{aslanpour2021autoscalesim} & 2021 & Java & \xmark & VMs, Cloudlets,web applications, end-user sessions & \cmark & Networking is abstracted and Static modeling limited to static thresholds and response time/delay metrics \\
			
			CloudSim-Express~\cite{hewage2024cloudsim} & 2023 & Java & \xmark & Data centers, VMs, cloudlets, power models & \xmark & No packet level network simulation and Scalability bound by cloudsim core \\
			
			CloudFactory~\cite{jacquet2023cloudfactory} & 2023 & Java & \xmark & VMs, workflows tasks, & \xmark & Limited to workflow-based applications and Stochastic modeling is predefined \\
			
			Cloudy~\cite{siavashi2024cloudy} & 2024 & Python & \xmark & CPU cores, RAM, GPU, V.RAM, V.CPU & \xmark & No Graphical interface and Lack of networking simulation \\
			
			\textbf{CloudyGUI} & \textbf{2025} & \textbf{Python} & \cmark & \textbf{CPU cores, RAM, GPU, Disk, V.RAM, V.CPU } & \cmark & \textbf{Lack of networking simulation} \\
			\bottomrule
		\end{longtable}
		
		\twocolumn
		
	\paragraph{Virtual Machine Monitor (VMM)}  
	The VMM is responsible for managing virtual machines on a given physical host. Its core functions include resource allocation, VM lifecycle management, and isolation between VMs. In the illustrated architecture, two VMMs (VMM1 and VMM2) are shown, each managing its own set of VMs.
	
	\paragraph{Virtual Machine (VM)}  
	A VM provides a virtualized computing environment. Each VM includes its own operating system kernel and hosts one or more containers. Containers are lightweight, isolated environments that run specific applications. In the diagram, VM1 runs Container A with App 1, while VM2 runs Container B with App 2.
	
	\paragraph{Resource Pool}  
	Located at the bottom of the DataCenter, the Resource Pool maintains a global view of available physical resources across all PMs.  It keeps track of CPU, memory, and GPU usage. It also supports dynamic resource provisioning. This pool is queried by the VM placement policy to inform allocation decisions.
	
	\paragraph{VM Placement Policy}  
	VM Placement Policy is located adjacent to the resource pool. It determines optimal VM placement based on current resource availability. The workflow is  as follows:
	\begin{itemize}
		\item The DataCenter receives a request to deploy a new VM.
		\item The VM Placement Policy consults the Resource Pool.
		\item Based on available resources, a suitable PM is selected.
		\item The corresponding VMM instantiates and manages the new VM.
		\item The VM runs its OS and hosts containers with applications.
		\item Resource usage is continuously monitored and managed by the VMM.
	\end{itemize}
	
	This hierarchical structure is well-suited for auto-scaling and often found in real-world cloud environments. This structure supports modular experimentation across various abstraction levels like hardware, virtualization and application. It is ideal for evaluation of auto scaling strategies.
	
		\subsection{Workload Generation}
		CloudyGUI uses a multi-stage pipeline to create cloud workloads. Initializing simulation parameters is the first step in the process, after which jobs and tasks are created and their interdependencies are modeled using a directed acyclic graph (DAG). 
		
		\subsubsection{Initialization Phase}
		The initialization phase establishes the fundamental parameters, resource constraints, and temporal scope for the entire simulation. It is a critical step that ensures the generated workloads are both realistic and feasible.
		
		\begin{enumerate}
			\item \textbf{Parameter Validation:}
			This step ensures the integrity of the user-defined inputs. The system validates three key parameters: Number of Jobs, Tasks per Job, and Instances per Task. It performs range checking to limit the values to a reasonable scale (e.g., 1-50,000 jobs, 1-20 tasks per job, 1-10 instances per task). This prevents the generation of an unmanageably large or computationally impossible workload. The system also performs type checking to ensure all inputs are valid, non-negative integers. This validation process is crucial for preventing errors and maintaining system stability.
			\item \textbf{Resource Pool Setup:}
			CloudyGUI initializes a default pool of system resources that represents the total available capacity for the simulation. This pool includes a fixed number of CPU cores (512), Memory (2TB), GPU units (32), and Disk space (10TB). The system monitors the consumption of these resources in real-time. Resource granularity is also defined, with memory tracked in MB and disk in GB which enables precise allocation and monitoring throughout the simulation.
			\item \textbf{Time Window:}
			A defined time window provides a temporal horizon for job scheduling, preventing all jobs from being submitted at the same time. CloudyGUI establishes a 7-day scheduling window starting from the current system time. This window serves as a reference for all job-related timestamps, such as submission and start times, and enables the simulation of realistic workload patterns over an extended period.
		\end{enumerate}

		\subsubsection{Job Generation Phase}\label{sec:jobgen}
		The job generation phase in CloudyGUI is a multi-step process that systematically builds a diverse and realistic workload from high-level job definitions to individual running instances.
		\begin{enumerate}
			\item \textbf{Job Type Selection:}
			CloudyGUI uses a weighted random selection process to choose job types, which ensures the generated workload reflects a realistic distribution of tasks in a cloud environment. For example, the system might assign a 30\% probability to a `data\_processing' job and a 25\% probability to a `machine\_learning' job. Each of these job types has predefined characteristics, including a specific duration range, a failure rate, and a set of typical tasks. The `web\_service' job type, for instance, has a shorter duration (15-120 minutes) and a low failure rate of 3\%, while the `machine\_learning' job type has a much longer duration (120-720 minutes) and a higher failure rate of 10\%.
			
			\item \textbf{Job Configuration:}
			Each generated job is assigned several key properties to define its behavior and place it within the simulation's time window. A unique ID is assigned, along with an initial status of `waiting'. The system sets a priority level from 1 (highest) to 5 (lowest). The job's duration is calculated randomly within the type-specific range, and a submission time is generated to distribute jobs realistically across the 7-day simulation window.
		
			\item \textbf{Task Creation:}
			For each job, CloudyGUI creates a number of tasks based on the user-defined tasks\_per\_job parameter. Each task is a smaller unit of work with its own unique ID, type, and resource requirements. These tasks inherit the priority from their parent job and have a duration proportional to the job's duration. The resource requirements, such as CPU, memory, GPU, and disk space, are determined based on the specific task type. For example, a `data\_preparation' task might require a specific range of CPU cores and a larger amount of memory compared to a lighter task.
		
			\item \textbf{Instance Generation:}
			Within each task, a number of instances are created, as defined by the instances\_per\_task parameter. Each instance is an individual execution unit with its own unique ID. The resources required for the task are scaled down and divided among its instances. For example, if a task needs 8 CPU cores and has 2 instances, each instance will be allocated 4 cores. The instance is assigned an initial status of `waiting', and its start and end times are determined during the simulation.
		
			\item \textbf{Status Management:}
			The system manages the status of each job, task, and instance as it progresses through the simulation, with statuses transitioning from `waiting' to `running' and finally to `completed', `failed', or `interrupted'.
	\end{enumerate}
	
	\subsubsection{Dependency Management}
	CloudyGUI's dependency management system is a comprehensive component of the workload generation process that ensures jobs are executed in the correct order. It reflects real-world interdependencies while preventing deadlocks. This process is crucial for generating realistic workload patterns that can be used to test and evaluate a scheduler's performance.
	
	\begin{enumerate}
		\item \textbf{Dependency Analysis:}
		The system establishes job dependencies based on predefined relationships between job types. For example, `machine\_learning' jobs may be defined to depend on `data\_processing' jobs. The system uses a weighted probability to determine actual job dependency, making the workload more dynamic. For instance, a `machine\_learning' job might have a 70\% chance of being dependent on a `data\_processing' job. The system also has the capability to handle complex relationships like circular dependencies.
		
		\item \textbf{Dependency Creation:}
		For each job that can have a dependency, the system identifies suitable parent jobs from the existing pool of jobs, ensuring that the parent job's submission time is earlier than the dependent job's. The system then constructs a directed acyclic graph (DAG) to represent the job dependencies. This graph is a crucial data structure for maintaining the order of operations and ensuring that jobs are not executed in an incorrect sequence.
		
		\item \textbf{Schedule Enforcement:}
		The system enforces dependencies by calculating the earliest possible start time for a job, which is determined by the completion time of all its parent jobs, including a small buffer time. CloudyGUI's scheduler is dependency-aware and only assigns resources to jobs once all their dependencies are satisfied. This mechanism ensures that the workload simulates a realistic workflow where tasks must be completed in a specific order. When a parent job completes, the scheduler checks if any dependent jobs can now be moved to the ready queue for execution.
		
		\item \textbf{Cycle Detection and Resolution:}
		To prevent deadlocks and logical errors, the system includes a safeguard to detect and resolve circular dependencies. It uses an algorithm (such as Depth-First Search) to check for cycles in the dependency graph. If a cycle is detected, the system automatically resolves it by removing a dependency to create a valid, acyclic graph, thus ensuring that a valid execution order for all jobs can be found.
		
		\item \textbf{Visualization and Monitoring:}
		CloudyGUI provides visualization tools to help users understand the complex relationships within the generated workload. The system can create visual representations of the dependency graph, which can be rendered as a PNG file. Additionally, the system tracks metrics related to dependencies, such as the average wait time of jobs due to dependencies, to provide a deeper analysis of the generated workload's characteristics.
	\end{enumerate}
	
	\subsubsection{Resource Allocation and Scheduling}
	
	CloudyGUI's resource allocation and scheduling system is designed to efficiently manage and distribute resources to jobs, mirroring the behavior of real-world cloud schedulers. This phase is crucial for simulating resource contention and evaluating the performance of different scheduling policies.
	
	\begin{enumerate}
		\item \textbf{Resource Tracking System:}
		The core of this system is the Resource Pool, which maintains a real-time view of available, total, and allocated resources. It uses thread-safe mechanisms to ensure accurate, concurrent updates. When a job needs resources, the system attempts to reserve them atomically. This process checks if the required resources are available and, if so, deducts them from the available pool and adds them to the allocated pool.
		
		\item \textbf{Scheduling Techniques:}
		CloudyGUI supports a range of scheduling techniques to manage job queues. A Priority-Based Scheduler organizes jobs into queues based on a 1-5 priority level, ensuring that higher-priority jobs are executed before lower-priority ones. To prevent potential issues like starvation where low-priority jobs could be indefinitely delayed by a continuous stream of high-priority tasks, the system implements an aging mechanism. This is achieved by having the scheduler periodically recalculate the effective priority of all jobs in the waiting queue at regular intervals (e.g., every 300 seconds). For every five minutes a job waits, its priority is boosted by one level (by subtracting from its numerical priority value). The priority queue is then rebuilt with these updated values, a process that guarantees fairness by ensuring that even low-priority jobs will eventually rise to the top of the queue and be executed. A Resource-Aware Scheduler adds another layer of intelligence by considering both a job's resource requirements and the current availability of those resources before making a scheduling decision. The system can also incorporate advanced techniques like preemption, where lower-priority jobs are temporarily suspended to make way for critical, high-priority jobs.
		
		\item \textbf{Resource Allocation Process:}
		The system employs an Admission Controller to manage the flow of jobs into the resource pool. When a job is submitted, the controller first checks if the necessary resources are available. If they are available, the resources are reserved, and the job's status is set to `running'. If not, the job is placed in a waiting queue.
		
		\item \textbf{Contention Handling:}
		The system employs a multi-faceted, proactive strategy to manage resource contention and optimize utilization. It uses a Resource Reservation System and Priority-Based Scheduling to guarantee that high-priority workloads have dedicated access to resources, even allowing preemption of lower-priority jobs. To prevent conflicts between different types of tasks, the system utilizes Resource Pooling and Workload Affinity, which map specific workloads to suitable hardware pools. 
		
		\item \textbf{Resource Monitoring and Adjustment:}
		CloudyGUI includes a real-time monitoring system that continuously tracks resource utilization. This system can be configured to automatically scale up or down resources based on predefined thresholds. For example, if CPU utilization exceeds 80\%, the system can simulate scaling up by adding more CPU cores to the resource pool. Conversely, if utilization drops below 20\%, it can scale down. The system can also address resource fragmentation, a state where a lack of contiguous resources prevents larger jobs from running, even if the total available resources are sufficient. This is handled by a defragmentation algorithm that might involve preempting and rescheduling jobs to consolidate resources.
		
		\item \textbf{Integration with Cloud Providers:}
		Finally, CloudyGUI's framework is designed to simulate the interaction with a real-world cloud provider. A dedicated Cloud Resource Manager can simulate scaling actions by requesting or releasing cloud resources. This module translates scaling decisions into API calls, simulating the provisioning of new virtual machine instances or the termination of existing ones, which is essential for evaluating auto-scaling strategies.
	\end{enumerate}
	
	\subsubsection{CSV Generation and Output Structure}
	
	CloudyGUI's CSV generation process is a robust data management system designed to capture a simulated workload in a structured format. This process ensures that the complex hierarchical data of jobs, tasks, and instances is flattened into a single file suitable for in-depth analysis.
	
	\begin{enumerate}
		\item \textbf{File Generation Process:}
		The process begins with an initialization step where the CSV file is created and a comprehensive set of headers is written. These headers cover information at the job, task, and instance levels, as well as additional details like dependencies. Following initialization, the system collects data from each component of the workload hierarchy. Job-level data, including ID, type, status, and resource requirements, is gathered first. This information is then integrated with data from its tasks and, subsequently, from each individual instance.
		
		\item \textbf{Hierarchical Data Flattening:}
		To translate the multi-level job structure into a flat CSV format, CloudyGUI employs a data flattening technique. It first collects a job's core data, then systematically adds task-level information to that data, and finally integrates instance-specific details. This results in each row of the CSV file representing a single instance, but containing all the contextual information from its parent task and job. This approach ensures that the relationships and properties of the entire workload are preserved in a single, accessible record.
		
		\item \textbf{CSV Writing Strategy:}
	    The system implements a reliable CSV writing strategy that ensures data integrity and handles concurrent access. It writes forecast data to a CSV file in the project's results directory using atomic writes with a temporary file that's renamed upon successful completion, preventing partial writes. The implementation automatically creates necessary output directories, verifies write permissions, and includes comprehensive error handling and logging. Timestamps are preserved as the index in the output CSV for time-series analysis. 
		
		\item \textbf{Data Validation and Analysis Integration:}
		The system ensures data integrity through post-processing steps that validate the generated CSV. This includes data type conversion, where fields are cast to their correct types, and data sanitization to handle missing or improperly formatted fields. The structured output is designed for seamless integration with data analysis tools. For example, the CSV file can be easily loaded into a Pandas DataFrame for advanced statistical analysis, allowing researchers to quickly calculate key metrics like total jobs, average job duration, and overall resource utilization. This integration bridges the gap between workload generation and actionable insights.
	\end{enumerate}
	
	\subsubsection{Post-Generation Processing}
	Post-generation processing in CloudyGUI is a critical phase for ensuring the validity and integrity of the simulated workload before it is used for analysis or testing. This phase performs a series of checks to confirm that the generated workload is within the defined system constraints.
	
	\begin{enumerate}
		\item \textbf{Workload Validation:}
		The system conducts a thorough validation of the generated workload to ensure it follows all predefined constraints. It performs System Constraints Verification to check if any job's resource requirements (e.g., CPU, memory) exceed the system's maximum limits. It also performs Temporal Validation to ensure that all time-based events, such as job start and end times, are logically sound and that no task starts before its parent job.
		
		\item \textbf{Resource Allocation Verification:}
		A key validation check is to prevent resource overallocation. The system processes a timeline of all job start and end events to calculate the total resource usage at every moment during the simulation. It then compares this peak usage against the system's total capacity for each resource (CPU, memory, GPU, etc.). This verification ensures that the generated workload does not demand more resources than the simulated system can physically provide at any given time, thus confirming the feasibility of the generated workload.

		\item \textbf{Dependency Management:}
		The system enforces job dependencies through a straightforward dependency tracking mechanism. Each job maintains a list of job IDs it depends on, which is validated during workload generation. The verifier ensures that no job starts before its dependencies complete by checking that all dependent jobs are in the 'terminated' state. The workload generator creates realistic dependency chains between jobs of different types while ensuring no circular dependencies are formed by only allowing dependencies on jobs created earlier in the sequence.
		
		\item \textbf{System Metrics Collection:}
		The system tracks and analyzes various metrics to evaluate prediction performance and system behavior. Key metrics include Mean Squared Error (MSE), Root Mean Squared Error (RMSE), Mean Absolute Error (MAE), R-squared ($R^2$), and Mean Absolute Percentage Error (MAPE). These metrics are calculated by comparing actual resource usage against predicted values, with special handling for edge cases like zero actual values to prevent division errors. The system also tracks explained variance to assess prediction quality. 
		
		\item \textbf{Summary Report and Visualization Generation:}
		CloudyGUI generates a detailed summary report that consolidates all validation results and key statistics into a single file. The system also automatically generates utilization plots using libraries like Matplotlib. For interactive analysis, key summaries and visualizations are rendered on the GUI's workload page using Chart.js. These visualizations show resource usage over time, with horizontal lines indicating the system's total capacity, which helps researchers visually identify periods of high and low utilization and spot potential bottlenecks. The generated report and plots provide the necessary insights to understand the workload and to inform the development of predictive and auto-scaling algorithms.
	\end{enumerate}
	
	\subsection{Resource Prediction}
	This section details the predictive modeling framework used in CloudyGUI to forecast resource demands. The system's ability to accurately predict future workload allows for proactive resource management, a key component of efficient auto-scaling.
	
	\begin{enumerate}
		\item \textbf{Data Collection and Preprocessing:}
		The system loads resource metrics from CSV files, focusing on numeric data columns and timestamps. The preprocessing pipeline includes several key steps: First, it identifies and parses timestamp columns, converting them to datetime objects. For data cleaning, it employs an IQR-based outlier detection method to identify and remove statistical outliers, replacing them with NaN values. While this statistical approach is robust for general noise, recent studies highlight the effectiveness of error distribution smoothing (EDS) for imbalanced regression in low-dimensional time series~\cite{CHEN2026115299}. This method offers a pathway to enhance prediction fidelity in volatile environments. Missing values are then handled through a three-step process: (1) linear interpolation between existing values, (2) forward-filling of any remaining NaNs, and (3) backward-filling of any remaining NaNs at the beginning of the dataset. The data is resampled to hourly intervals using mean aggregation, and the system includes comprehensive logging at each preprocessing step to track data quality issues and processing outcomes.
		
		\item \textbf{Feature Engineering:}
		To enhance prediction accuracy, a comprehensive set of features is engineered from the preprocessed time-series data. This includes:
		\begin{itemize}
			\item \textbf{Time-based Features:} Extraction of temporal attributes such as the hour of the day, day of the week, and weekend flags to capture cyclical patterns in workload behavior.
			\item \textbf{Statistical Features:} Calculation of rolling statistics (mean, standard deviation, and maximum) over various time windows (e.g., 5, 15, and 60 minutes) to capture short-term trends and volatility.
		\end{itemize}
		\item \textbf{Data Leakage Prevention:}
			To strictly prevent look-ahead bias\footnote{Look-ahead bias occurs when future information is inadvertently used in model training or validation, leading to overly optimistic performance estimates.} and to ensure that our $R^{2}$ values reflect genuine predictive power, we implemented a robust pipeline that enforces temporal causality. We utilize a \texttt{time\_series\_train\_test\_split} strategy where the data split occurs before any feature engineering. Specifically, we adhere to the following protocols:
			\begin{enumerate}
				\item \textbf{Isolated Transformation:} Feature scaling is performed using a \texttt{fit\_transform} operation exclusively on the training partition. The test partition is subsequently processed using \texttt{transform} with the training set's statistics, ensuring that global distribution metrics (mean, variance) from the future do not leak into the training process.
				\item \textbf{Safe Feature Generation:} Rolling statistics (e.g., 3-hour mean) are calculated using a fixed window on shifted data ($\text{Lag}_{t-1}$) within each fold. This guarantees that the feature vector at any time $t$ is derived solely from historical data ($t-k$ to $t-1$).
				\item \textbf{Strict Chronological Validation:} We employ \texttt{TimeSeriesSplit} for cross-validation, which respects temporal ordering by ensuring that each training fold only accesses data preceding the validation index. This strictly avoids the use of future information during interpolation or resampling.
		\end{enumerate}

		\item \textbf{Model Architecture and Training:}
		The framework uses various predictive models, including gradient boosting regressors such as XGBoost, and deep learning models such as Long Short-Term Memory (LSTM) networks. Since XGBoost has shown strong performance in our preliminary comparative analysis and is well-suited for time-series forecasting \cite{bawa2025improving}, we chose to focus on it for our primary predictions. Specifically, our own previous work demonstrated the efficacy of XGBoost in workload prediction, where it achieved superior results (e.g., an $R^2$ value of $0.97967$ for CPU utilization) compared to baseline models. The training pipeline uses a chronological data split to simulate a realistic scenario where models are trained on past data and evaluated on entirely unseen future data. Specifically, for a given 7-day workload, the initial 80\% of the data (approximately 5.6 days) is allocated for training, while the final 20\% (approximately 1.4 days) is reserved for testing. The training process incorporates callbacks like early stopping to prevent overfitting and ensure the models generalize well.

		\item \textbf{Resource Prediction:}
		Formally, the resource prediction is modeled as a supervised regression task where the objective is to map an input feature vector $X_t$ at time $t$ to a predicted resource utilization $\hat{y}_{t+1}$ (e.g., CPU, Memory). Based on our feature engineering pipeline, the input vector $X_t$ is defined as:

				\begin{equation}
					X_t = [\mathbf{T}_t, \text{Lag}_{1h}(y), \text{Lag}_{24h}(y), \mu_{3h}(y)]
				\end{equation}
				
				\noindent where:
				\begin{itemize}
					\item $\mathbf{T}_t$ represents the temporal features: $\{\text{hour}, \text{day\_of\_week}, \text{day\_of\_year}, \text{month}\}$.
					\item $\text{Lag}_{k}(y)$ denotes the historical utilization values at time $t-k$.
					\item $\mu_{3h}(y)$ is the 3-hour rolling mean used to capture short-term trends.
				\end{itemize}
				
				For the primary XGBoost model, the learning objective is to minimize the regularized squared error loss:
				
				\begin{equation}
					\mathcal{L}(\phi) = \sum_{i} (y_i - \hat{y}_i)^2 + \sum_{k} \Omega(f_k)
				\end{equation}
			
		\noindent Here, the first term represents the Mean Squared Error (MSE) between the actual ($y_i$) and predicted ($\hat{y}_i$) usage, and $\Omega(f_k)$ serves as the regularization term to penalize tree complexity and prevent overfitting. The model was trained with $n\_estimators=100$ and a learning rate of $0.1$.

		\item \textbf{Model Evaluation and Monitoring:}
		The performance of the predictive models is evaluated using standard regression metrics, including Mean Absolute Error (MAE), Root Mean Squared Error (RMSE), and the coefficient of determination ($R^2$).
	\end{enumerate}

	\subsection{Predictive Threshold-Based Auto-scaling}
	
	This system has an automated control mechanism that uses predicted future resource usage to proactively adjust cloud resources. Instead of reacting to a problem that has already occurred, it anticipates future needs. The entire process is structured around a \emph{MAPE} control loop, a fundamental framework for autonomous systems.
	
	\begin{enumerate}
		\item \textbf{Core Components:}
		The system's architecture consists of several integrated services that correspond directly to the MAPE loop phases:
		\begin{itemize}
			\item \textbf{Monitor Service:} This component handles the \emph{Monitor} phase, continuously gathering real-time resource metrics like CPU, memory, and network I/O.
			\item \textbf{Prediction Engine:} This is the core of the \emph{Analyze} phase, using forecasting models (e.g., XGBoost, LSTM) to predict future resource needs based on the data collected by the monitor.
			\item \textbf{Decision Engine:} This component executes the \emph{Plan} phase. It applies scaling rules and thresholds to the predictions to determine if an action is needed. This logic uses a decision matrix to handle various scenarios, such as making an aggressive scale-up if a critical threshold is predicted to be breached.
			\item \textbf{Executor Service:} This service performs the \emph{Execute} phase, carrying out the planned scaling actions by communicating with the underlying cloud provider to provision or de-provision resources.
		\end{itemize}
		
		\item \textbf{Scaling Logic:}
		The system uses a predictive threshold-based approach to make decisions. Thresholds have predefined upper and lower utilization limits, while the Prediction Engine forecasts resource usage for future time windows. The Decision Engine then analyzes these predictions against the thresholds to determine the necessary action. This scaling approach aligns with recent deep learning-based auto-scaling frameworks for Service Function Chains in cloud environments~\cite{taha2024proactive}.

		\item \textbf{Implementation and Advanced Features:}
		The implementation includes features like adaptive thresholds, which automatically adjust based on historical patterns, and cost optimization to balance performance with cost. The system continuously evaluates its own performance through model monitoring and can retrain models if their prediction accuracy degrades, ensuring the Analyze phase remains reliable. The framework is also designed to be extensible, with support for different cloud providers for the Execute phase.
	\end{enumerate}
	Overall, the system provides an efficient and reliable way to manage cloud resources by using predictive intelligence within a self-managing MAPE loop to maintain performance and control costs.
	
	\subsection{Simulated Auto-scaling}
	The predictive threshold-based system demonstrated the feasibility of proactive resource management using real-time metrics and forecasting models. However, its reliance on actual cloud providers introduced constraints in terms of cost, repeatability, and experimental control. To overcome these limitations and enable rigorous evaluation of scaling policies under diverse workload scenarios, we transitioned to a fully simulated environment. This simulation framework abstracts the behavior of cloud infrastructure and allows deterministic experimentation with auto-scaling logic. It preserves the core principles of the MAPE loop by monitoring synthetic workloads, analyzing resource metrics, and planning scaling actions. These actions are then executed within a controlled virtual setup, offering fine-grained control over timing, state transitions, and feedback loops. The following components collectively form the backbone of this simulated auto-scaling system.
	
	\begin{enumerate}
		\item \textbf{SimulationRunner:} The SimulationRunner is the central controller that governs the entire simulation process. Its primary role is to manage the simulation's lifecycle, timing, and data aggregation.
		\begin{itemize}
			\item \textbf{Lifecycle and Execution Flow Management:} The runner initiates the simulation based on a given configuration file. It is responsible for the setup (initializing the mock cloud provider and scaling engine), execution (running the main simulation loop for a predetermined duration), and termination. It controls the simulation's clock, advancing time in discrete steps. At each time step (e.g., every 30 seconds), it directs the sequence of events: triggering the workload generator, instructing the mock cloud provider to update its instance metrics, and invoking the scaling engine to evaluate the current state. This precise control over the execution flow ensures that experiments are deterministic and reproducible, which is critical for academic research.
			
			\item \textbf{Data Collection and Logging:} Throughout the simulation, the SimulationRunner serves as the primary data logger. It systematically records every event and state change, including the generated workload intensity, the resource metrics reported by the cloud provider (CPU, memory usage), the decisions made by the scaling engine, and the number of active instances. This comprehensive data collection is foundational for the post-simulation analysis, as it provides the raw material for generating plots, calculating performance statistics, and validating the effectiveness of the tested scaling policy.
			
	  		\item \textbf{Resource Usage Mapping:} To ensure high-fidelity simulation, CloudyGUI maps abstract workload tasks to concrete resource utilization metrics (CPU, Memory, Disk, GPU) using a profile-based dynamic model. As defined in the \texttt{TASK\_TYPES} configuration in Section~\ref{sec:jobgen}, distinct workload categories exhibit unique resource footprints. For example, \textit{Model Training} tasks are CPU/GPU intensive (utilizing 70--100\% CPU and 80--100\% GPU), whereas \textit{Data Ingestion} tasks are I/O bound (utilizing 70--90\% Disk).

			The real-time utilization $U_{r}(t)$ for a specific resource $r$ during a task's execution is calculated dynamically based on its progress and a stochastic noise factor. The mapping function is defined as:
			
			\[
				U_{r}(t) = C_{\text{req}} \times \left( \alpha_{\text{min}} + (\alpha_{\text{max}} - \alpha_{\text{min}}) \times \frac{t_{\text{elapsed}}}{T_{\text{duration}}} + \delta \right)
				\]
		where:
			\begin{itemize}
				\item $C_{\text{req}}$ is the allocated resource capacity (e.g., 4 vCPUs).
				\item $[\alpha_{\text{min}}, \alpha_{\text{max}}]$ is the resource pattern range defined for the specific task type (e.g., $[0.4, 0.8]$ for data ingestion).
				\item $\frac{t_{\text{elapsed}}}{T_{\text{duration}}}$ represents the task completion progress (0.0 to 1.0).
				\item $\delta \sim \text{Uniform}(-0.05, 0.05)$ introduces stochastic fluctuation to mimic real-world volatility.
			\end{itemize}	
			This mapping ensures that resource metrics accurately reflect the lifecycle of the underlying tasks, including ramp-up phases and varying intensity, rather than static allocation.	
		\end{itemize}
		
		\item \textbf{MockCloudProvider:} The MockCloudProvider is a high-fidelity abstraction of a real-world cloud infrastructure provider (like AWS EC2 or Google Compute Engine). Its purpose is to create a realistic, yet fully controlled and cost-free, environment for the scaling engine to operate within.
		\begin{itemize}
			\item \textbf{Infrastructure and State Simulation:} This component simulates the core behaviors of a cloud environment. When the scaling engine requests a new instance, the MockCloudProvider does not provision a physical server. Instead, it simulates the process by creating a virtual instance object and transitioning it through realistic states: from \texttt{pending} to \texttt{running}. This includes modeling the inherent delays associated with instance boot-up times. Similarly, when an instance is terminated, it moves to a \texttt{terminating} state before being removed. This state tracking is crucial because it accurately models the real-world lag between a scaling decision and its effect on the system's capacity.
			
			\item \textbf{Dynamic Resource Metrics Provision:} The most critical function of the MockCloudProvider is to supply the scaling engine with realistic resource metrics. These metrics are not static; they are dynamically calculated at each time step based on the current synthetic workload and the number of instances in a \texttt{running} state. For example, if the workload increases while the instance count remains the same, the provider will report a higher average CPU utilization. This creates the essential feedback loop for the auto-scaling logic: the environment's state changes in response to the workload, and the scaling engine reacts to those changes.
		\end{itemize}
		
		\item \textbf{EnhancedScalingEngine:}  The EnhancedScalingEngine is the brain of the auto-scaling system. It encapsulates the logic and policies that are the subject of the research. Its sole responsibility is to analyze the state of the system and make intelligent scaling decisions.
		\begin{itemize}
			\item \textbf{Implementation of Scaling Policies:} This is where the core scaling algorithm is implemented. In its basic form, the policy is based on thresholds: if the average resource usage exceeds a \texttt{scale\_up\_threshold} (e.g., 70\%), the engine decides to add instances. If usage falls below a \texttt{scale\_down\_threshold} (e.g., 40\%), it decides to remove instances. However, this engine is designed to be "enhanced," meaning researchers can implement more sophisticated policies, such as predictive algorithms that analyze trends in resource usage to scale proactively, or multi-metric policies that consider CPU, memory, and network latency simultaneously.
			
			\item \textbf{Constraint Handling:} Beyond simple decision-making, the engine is responsible for enforcing real-world operational constraints. The most important of these is the cooldown period. After a scaling action is initiated, the engine enters a cooldown phase during which it will not make further scaling decisions. This prevents system instability known as "thrashing", where the system rapidly scales up and down in response to short-term metric fluctuations. By handling these constraints, the engine ensures that its decisions are not just logically sound but also practical and safe for a production environment.
		\end{itemize}
		
	\end{enumerate}
	
	In summary, CloudyGUI provides a modular and controllable environment for simulating auto-scaling strategies under varied conditions. Its integration of realistic workload generation, predictive modeling, and a structured simulation loop supports reproducible experimentation and policy evaluation. 
	
	\section{Results and Analysis}\label{sec4}
	This section presents a detailed analysis of the simulation results generated by CloudyGUI, demonstrating the framework's effectiveness in modeling and managing dynamic cloud workloads. The findings are organized into three subsections that directly correspond to our core methodology: an evaluation of the Workload Generation component, an analysis of the Resource Prediction model's accuracy, and a demonstration of the Predictive Auto-scaling system's performance.
	\subsection{Workload Generation}
	This section analyzes the characteristics of the workload produced by the CloudyGUI framework to validate its complexity and realism. The generated workload is composed of a diverse set of jobs and tasks, designed to rigorously test the performance of cloud scheduling and auto-scaling
	algorithms.
	
	To ensure a comprehensive output, the generated workload is captured in a structured format, the metadata structure of which is detailed in Table~\ref{tab:job_instance_metadata}. Organized hierarchically, each row in the final output file represents a single instance while retaining the full context of its parent task and job. This structure is designed to capture all essential attributes for evaluation, including unique identifiers for tracing (e.g., Job ID, Instance ID), scheduling parameters like priority and dependencies, and the crucial distinction between requested resources (e.g., CPU Required) and actual measured usage (e.g., CPU Usage). This multi-level dataset provides the necessary granularity for in-depth analysis of system behavior, scheduler performance, and the effectiveness of resource management strategies.

\begin{figure}
		%\centering
		\includegraphics[width=\linewidth]{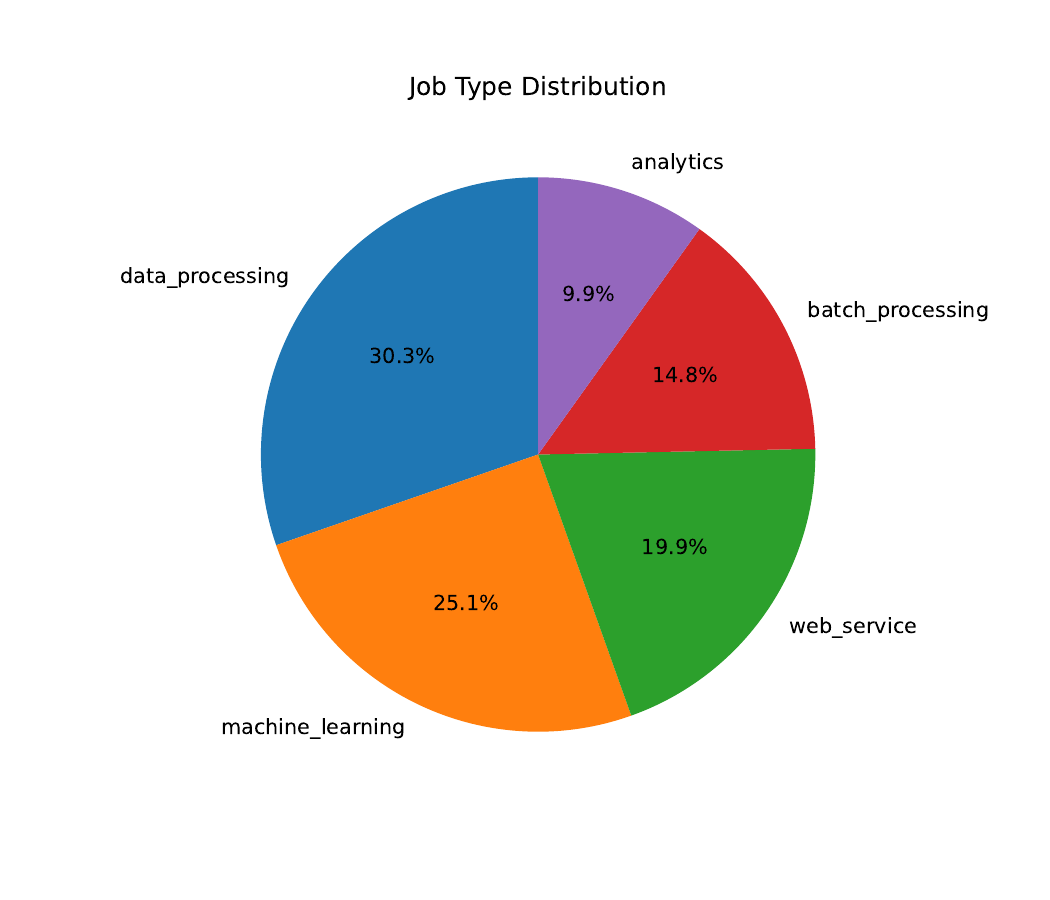}
		\caption{Distribution of Jobs}
		\label{fig:jobtypedistribution}
	\end{figure}

\begin{figure}[!h]
		%\centering
		\includegraphics[width=\linewidth]{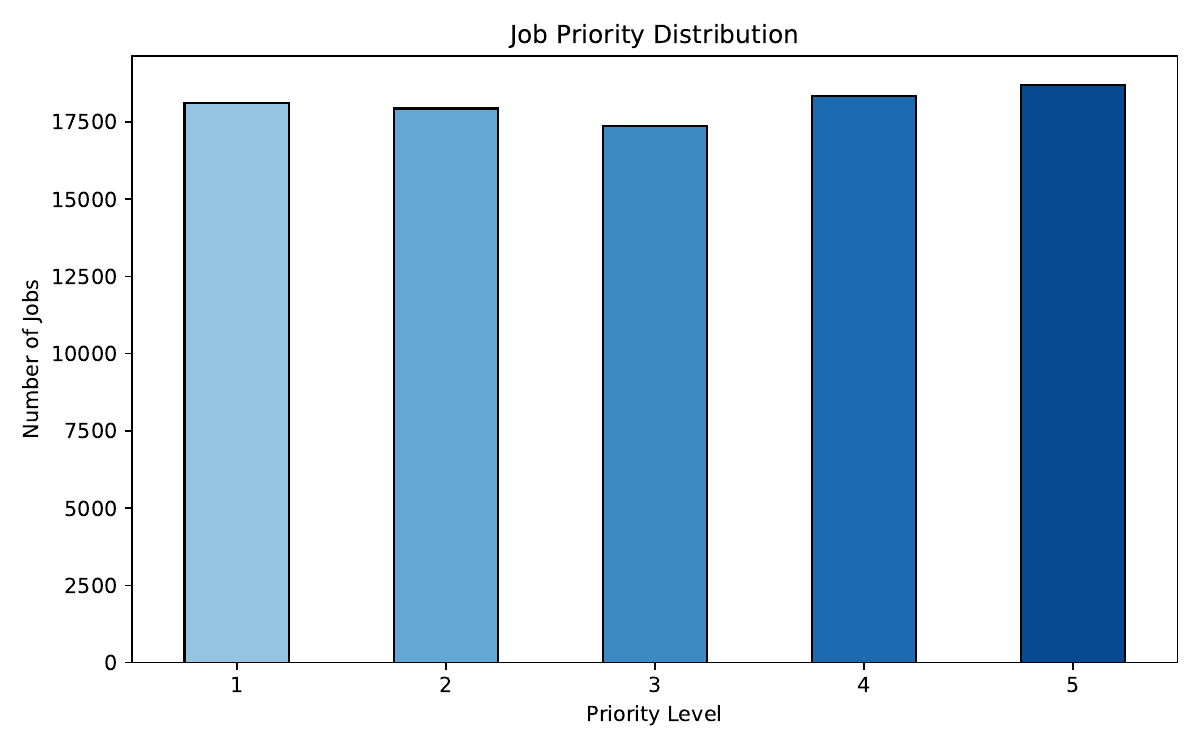}
		\caption{Priority Distribution of jobs}
		\label{fig:jobprioritydistribution}
	\end{figure}

	We configured the workload generator with 10,000 jobs, 5 tasks per job, and 5 instances per task to derive the following graphs. Figure \ref{fig:jobtypedistribution} illustrates the proportional distribution of job types within the generated workload. The composition is diverse, with a significant share of 'data\_processing' (30.3\%) and 'machine\_learning' (25.1\%) jobs, reflecting common, resource-intensive cloud use cases. This heterogeneity is a direct result of the weighted selection process and ensures the simulation is representative of a varied cloud environment. Figure \ref{fig:jobprioritydistribution} displays the distribution of job priorities. The number of jobs is nearly uniform across all five priority levels, which is essential for an unbiased evaluation of priority-aware scheduling policies by ensuring a balanced contention scenario between high and low-priority tasks.
	
	\begin{table*}[!h]
		\centering
		\caption{Hierarchical Metadata Schema for Generated Workloads}
		\begin{tabular}{|l|p{10cm}|}
			\hline
			\textbf{Attribute} & \textbf{Description} \\
			\hline
			\multicolumn{2}{|c|}{\textbf{Job Information}} \\
			\hline
			Job ID & Unique identifier for each job. \\
			Job Type & Category or classification of the job. \\
			Task ID & Unique identifier for each task within a job. \\
			Task Type & Category of the task. \\
			Priority & Priority level of the job/task. \\
			\hline
			\multicolumn{2}{|c|}{\textbf{Dependencies}} \\
			\hline
			Dependencies & List of task IDs this task depends on. \\
			Dependency Types & Types of dependencies between tasks. \\
			\hline
			\multicolumn{2}{|c|}{\textbf{Instance Information}} \\
			\hline
			Instance ID & Unique identifier for the compute instance. \\
			Instance Status & Current state of the instance (e.g., running, completed, failed). \\
			Instance Start Time & When the instance started processing. \\
			Instance End Time & When the instance finished processing. \\
			\hline
			\multicolumn{2}{|c|}{\textbf{Resource Requirements}} \\
			\hline
			CPU Required & Number of CPU cores requested. \\
			CPU Usage & Actual CPU cores used. \\
			Memory Required (MB) & Memory requested in MB. \\
			Memory Usage (MB) & Actual memory used in MB. \\
			GPU Required & Number of GPUs requested. \\
			GPU Usage & Actual GPUs used. \\
			Disk Required (GB) & Disk space requested in GB. \\
			Disk Usage (GB) & Actual disk space used. \\
			\hline
			\multicolumn{2}{|c|}{\textbf{Virtual Machine Details}} \\
			\hline
			VM ID & Identifier for the virtual machine. \\
			VM CPU & Total CPU cores available on the VM. \\
			VM RAM & Total RAM available on the VM in MB. \\
			VM GPU & Total GPUs available on the VM. \\
			\hline
		\end{tabular}
		\label{tab:job_instance_metadata}
	\end{table*}
	
	\begin{figure*}[!h]
		\centering
		\includegraphics[width=\linewidth]{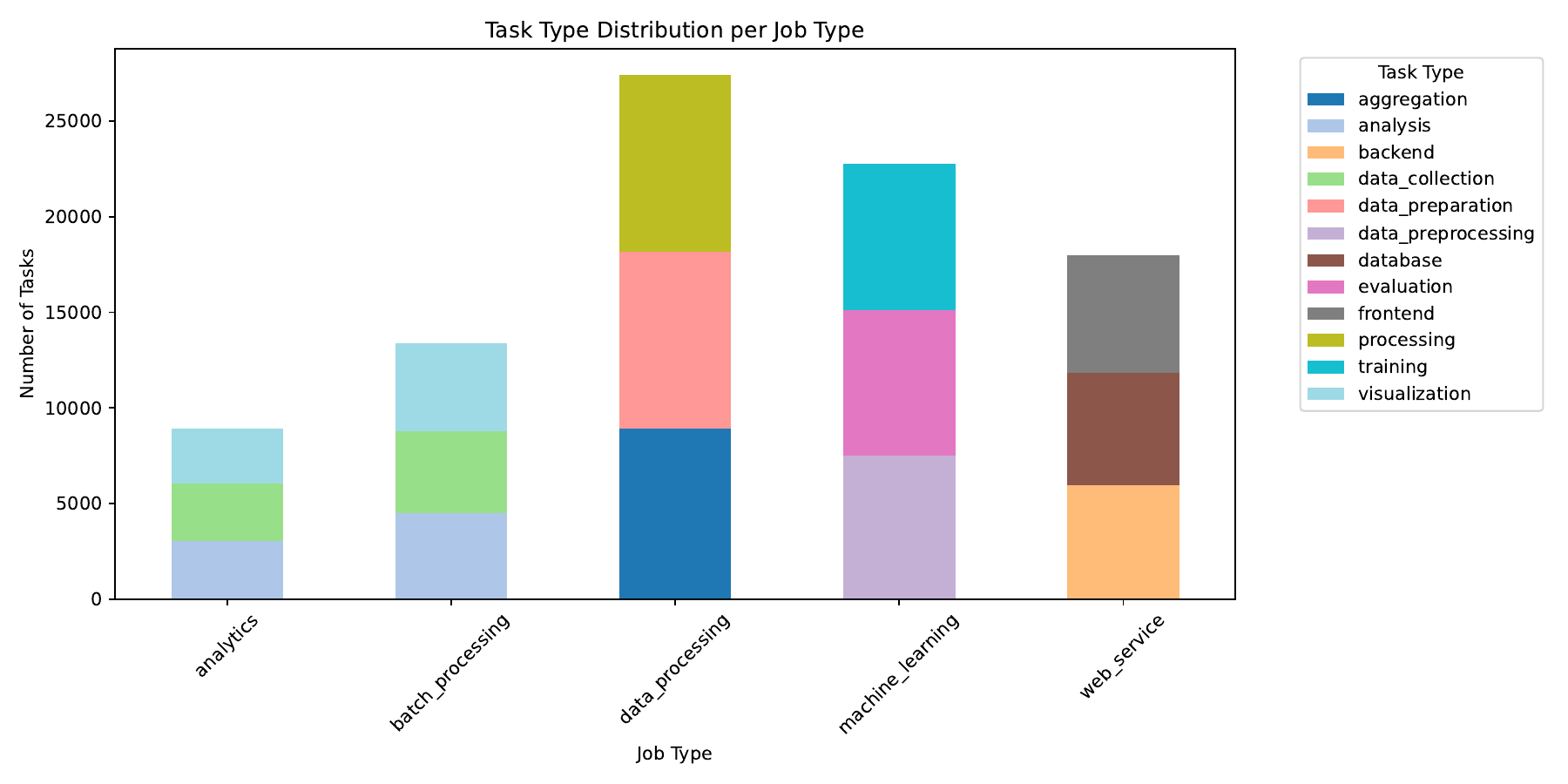}
		\caption{Compositional breakdown of task types per job}
		\label{fig:tasktypedistributionperjobtype}
	\end{figure*}
		\begin{figure}[!h]
		\centering
		\includegraphics[width=\linewidth]{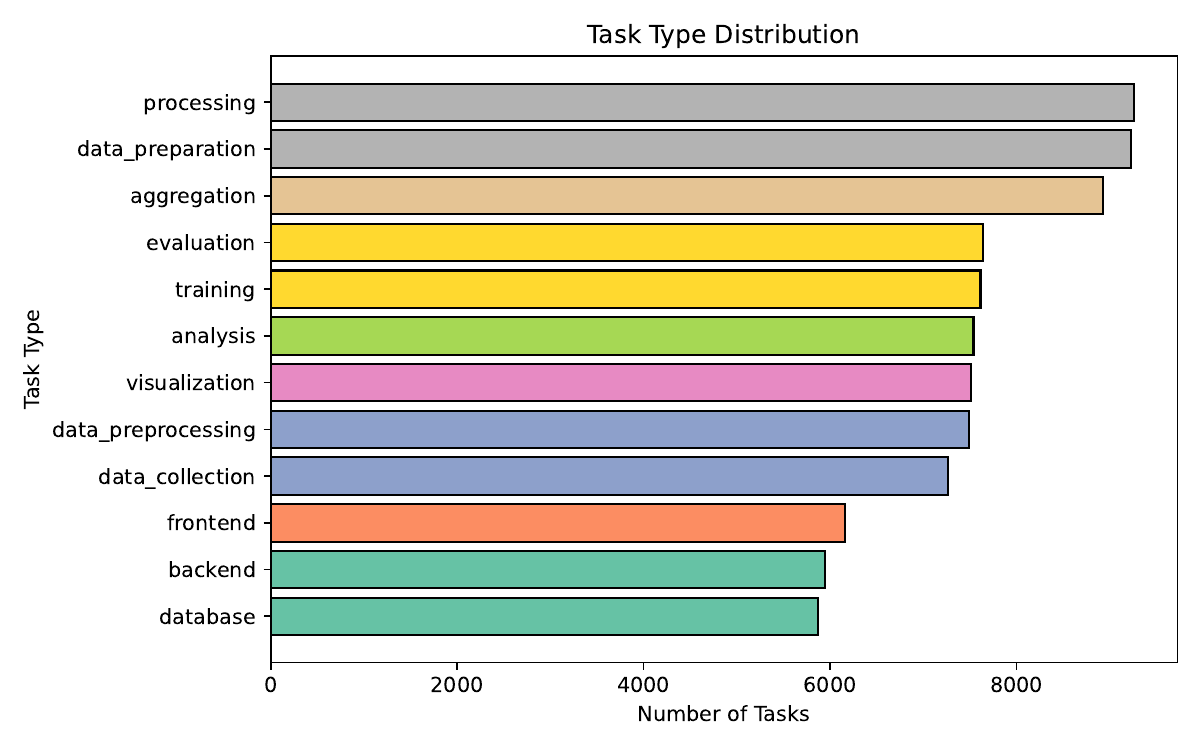}
		\caption{Aggregate frequency of each task type across the entire workload}
		\label{fig:tasktypedistributionhorizontalsorteddesc}
	\end{figure}
	
	Figure \ref{fig:tasktypedistributionperjobtype} confirms the structural integrity of the generated jobs by providing a breakdown of the task types that constitute each job type. For instance, it shows that 'machine\_learning' jobs are composed of tasks like 'data\_preprocessing', 'training', and 'evaluation'. The aggregate frequency of each task type across the entire workload is shown in 
	
\begin{figure}[!h]
		\centering
		\includegraphics[width=\linewidth]{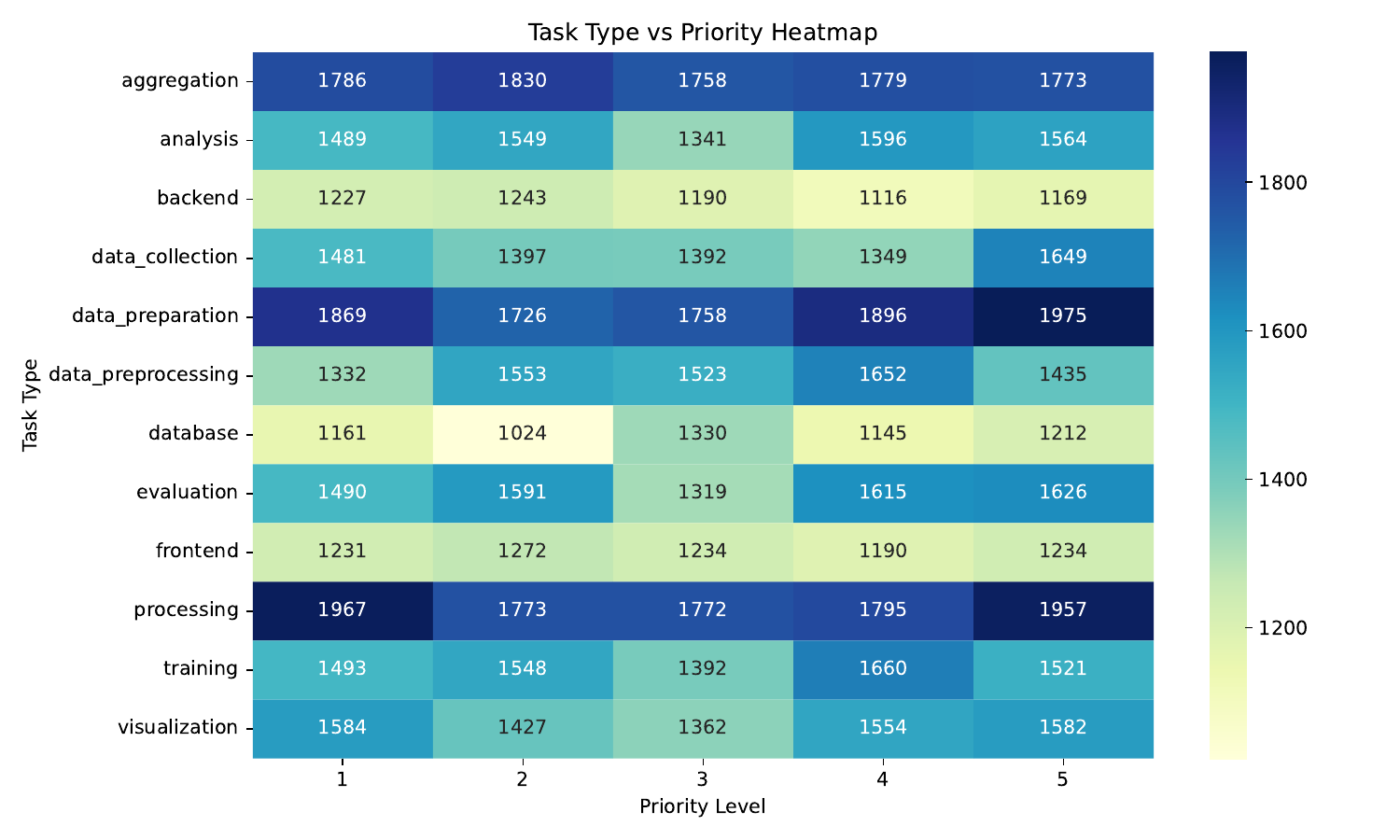}
		\caption{Heatmap visualizing the relationship between task types and Priorities}
		\label{fig:tasktypeheatmap}
	\end{figure}

	Figure \ref{fig:tasktypedistributionhorizontalsorteddesc}, where 'processing' and 'data\_preparation' are the most common, underscoring the data-intensive nature of the simulation. Finally, the heatmap in Figure \ref{fig:tasktypeheatmap} visualizes the relationship between task types and their assigned priorities. The data confirms that priorities are well-distributed across all task types, preventing any single category from being systematically favored and thus creating a challenging and realistic scenario for evaluating advanced scheduling algorithms.

	\begin{figure}[t]
		\centering
		\includegraphics[width=\linewidth]{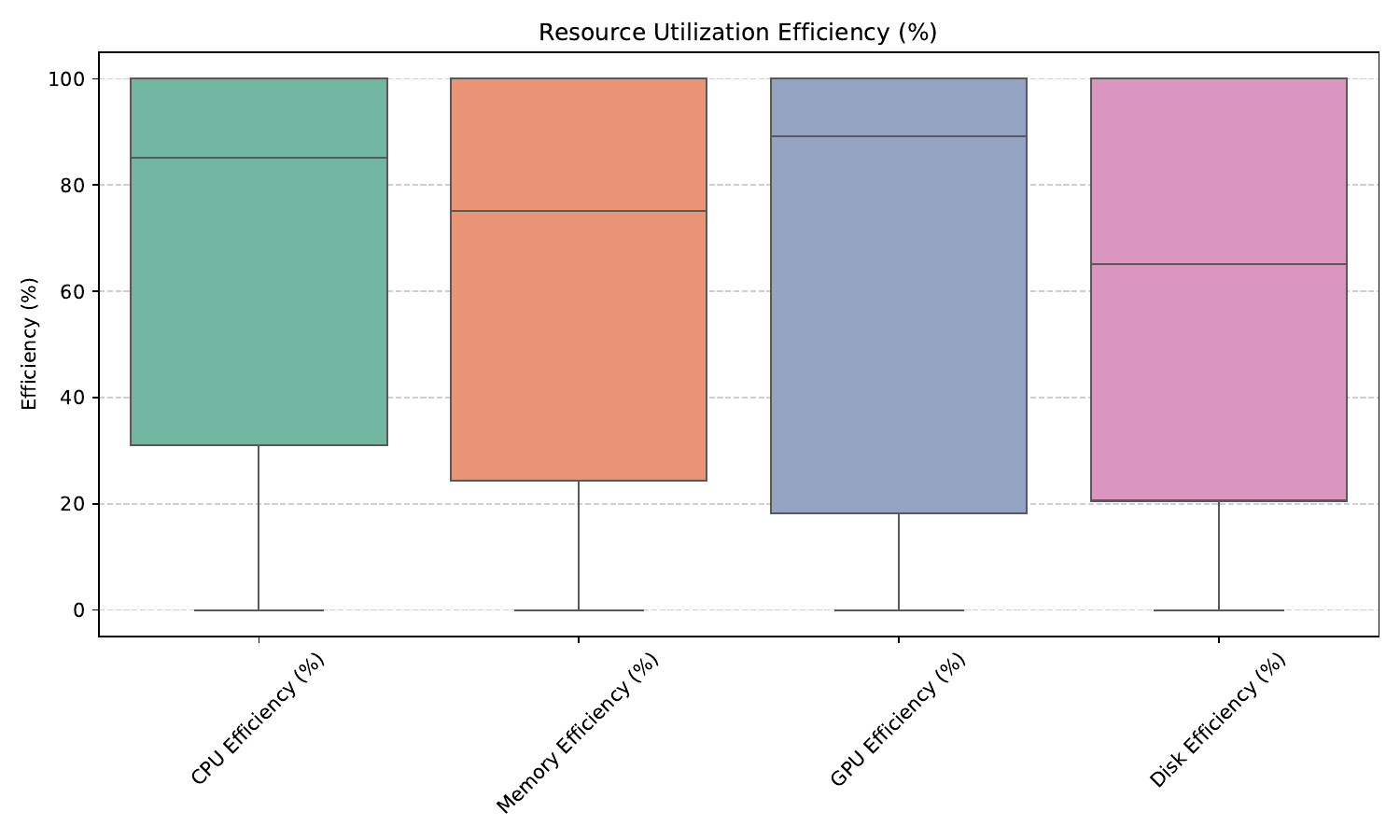}
		\caption{Box plots of resource utilization}
		\label{fig:resourceutilization}
	\end{figure}
	\begin{figure}
		\centering
		\includegraphics[width=\linewidth]{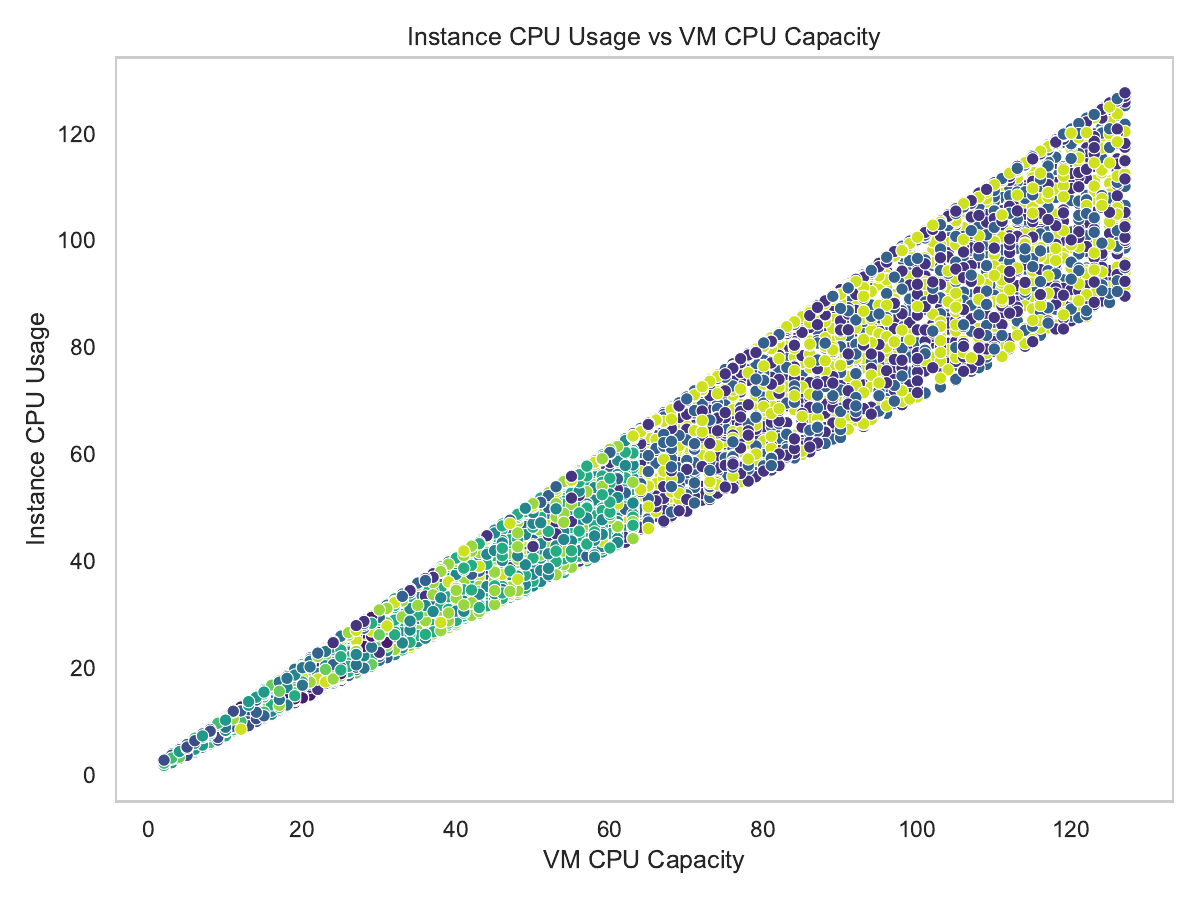}
		\caption{Instance CPU usage versus VM capacity}
		\label{fig:instancecpuusagevsvmcapacity}
	\end{figure}
	
	Furthermore, the efficiency of resource consumption is summarized in Figure \ref{fig:resourceutilization}. The median utilization for CPU and Memory was approximately 85\% and 75\%, respectively, with significant variance indicating periods of both high and low usage.
	
	Figure \ref{fig:instancecpuusagevsvmcapacity} validates the simulation's physical constraints by plotting instance CPU usage against the host VMs CPU capacity. As expected, all data points fall on or below the main diagonal, confirming that no instance exceeded the resources assigned to its VM.

	\begin{figure}[!h]
		\centering
		\includegraphics[width=\linewidth]{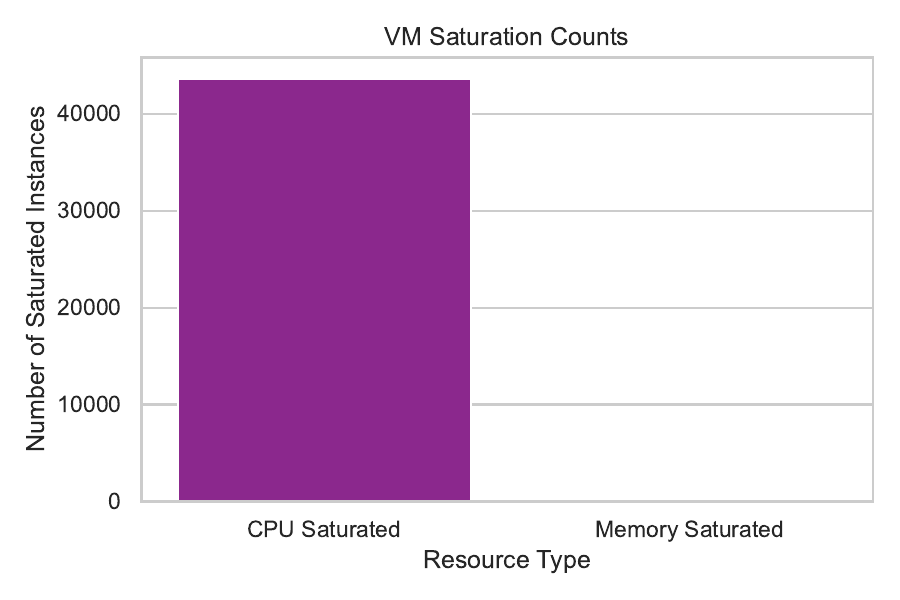}
		\caption{Counts of resource saturation events}
		\label{fig:vmsaturationcounts}
	\end{figure}
	
	 Figure \ref{fig:vmsaturationcounts}, quantifies resource saturation events and the results show that over 40,000 instances became CPU-saturated, while none reached memory saturation. This identifies the workload as heavily CPU-bound, meaning processing power is the primary limiting resource, a crucial characteristic for testing resource provisioning strategies.
	
\begin{figure}[!h]
		\centering
		\includegraphics[width=\linewidth]{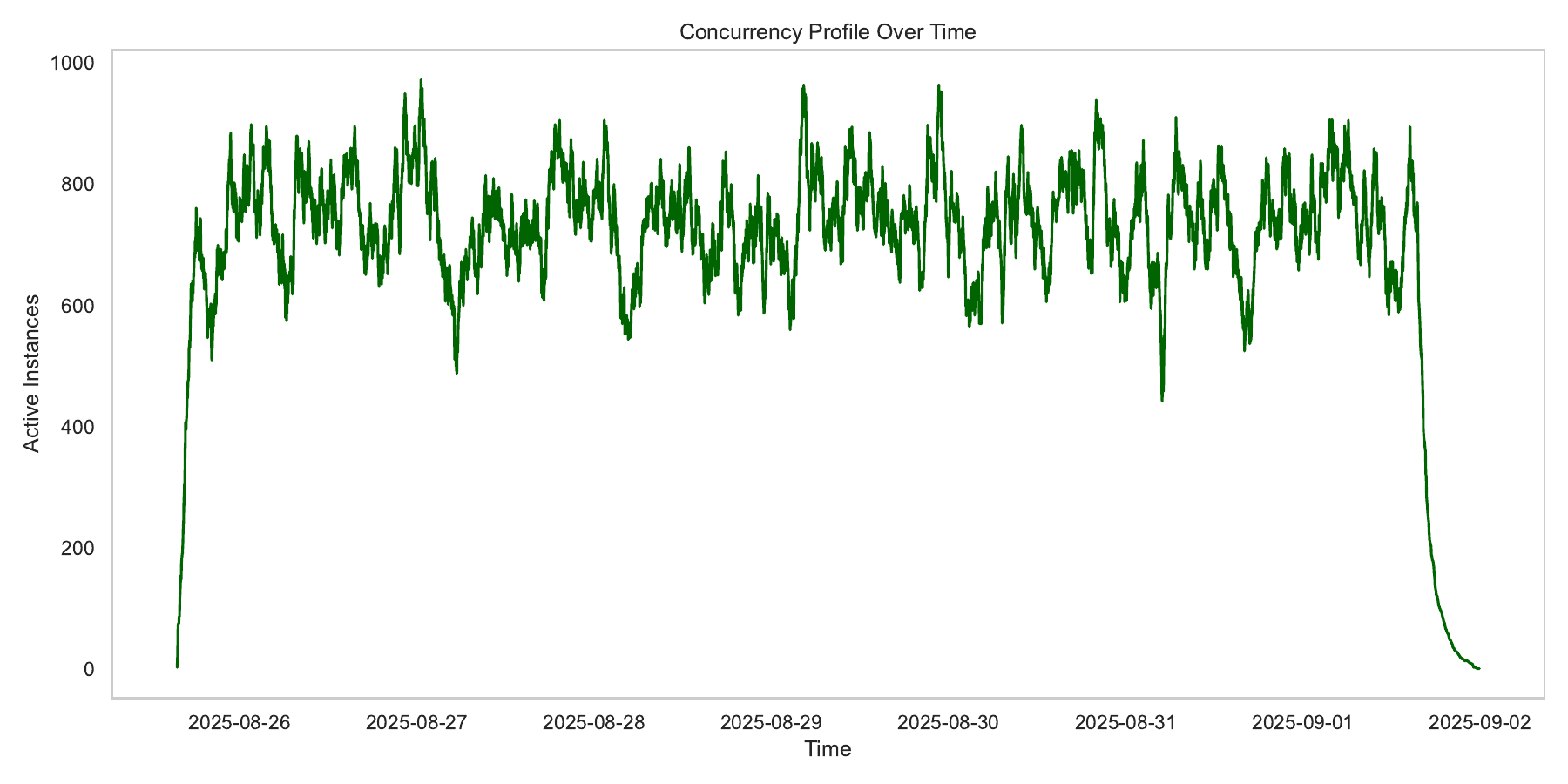}
		\caption{Concurrency profile over a 7-day period}
		\label{fig:concurrencyprofileovertime}
	\end{figure}

	Finally, Figure \ref{fig:concurrencyprofileovertime} presents the concurrency profile over the 7-day simulation period, plotting the number of active instances over time. The plot reveals a highly dynamic environment, with concurrent instances fluctuating between approximately 500 and 900. This non-static, fluctuating demand profile confirms that the generated workload provides a realistic and challenging testbed for evaluating the responsiveness and efficacy of auto-scaling systems.
	%\clearpage
	%\newpage
	
	\subsection{Resource Prediction}
	The resource prediction phase is a critical component of CloudyGUI's proactive auto-scaling system, representing the "Analyze" stage of the MAPE loop. This section presents a visual analysis of the predictive model's performance, showcasing its ability to forecast future resource utilization across different resource types. As a single-step forecasting model, the system is designed to predict resource usage in the near future. The performance is evaluated by observing how closely the predicted values track the actual resource usage over time.
	
	\begin{table*}[htbp]
		\centering
		\caption{Prediction Results Summary}
		\begin{tabular}{lrrrrr}
			\toprule
			\textbf{Resource} & \textbf{RMSE} & \textbf{MSE} & \textbf{MAE} & \textbf{R\textsuperscript{2}} & \textbf{EVS} \\
			\midrule
			GPU    & 0.062659  & 0.003926   & 0.024194  & 0.997014  & 0.997016 \\
			CPU    & 2.773768  & 7.693787   & 1.634416  & 0.986673  & 0.986682 \\
			Memory & 12.073929 & 145.779770 & 7.214408  & 0.983097  & 0.983097 \\
			Disk   & 70.037719 & 4905.282013& 45.585369 & 0.962239  & 0.962262 \\
			\bottomrule
		\end{tabular}
		\label{tab:prediction_summary}
	\end{table*}
	
\begin{figure*}[!h]
		\centering
		\includegraphics[width=\linewidth]{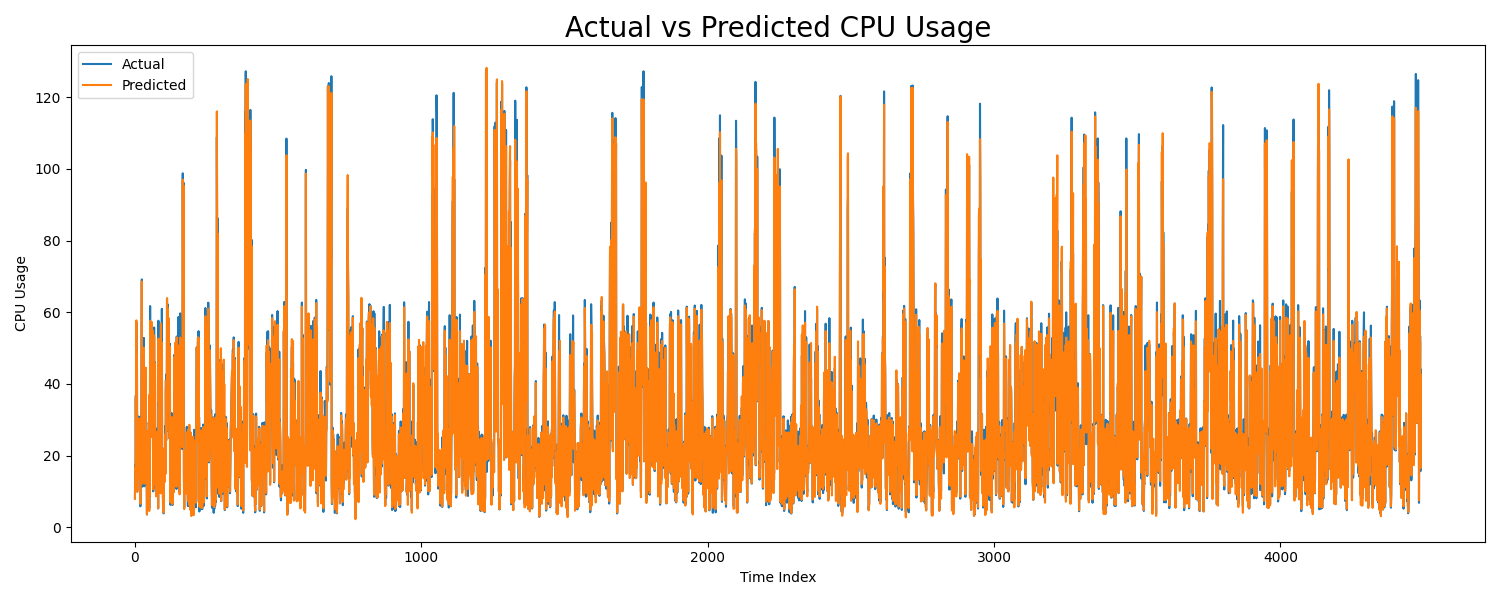}
		\caption{Comparison of actual versus predicted CPU usage}
		\label{fig:cpuprediction}
	\end{figure*}

\begin{figure*}[!h]
		\centering
		\includegraphics[width=\linewidth]{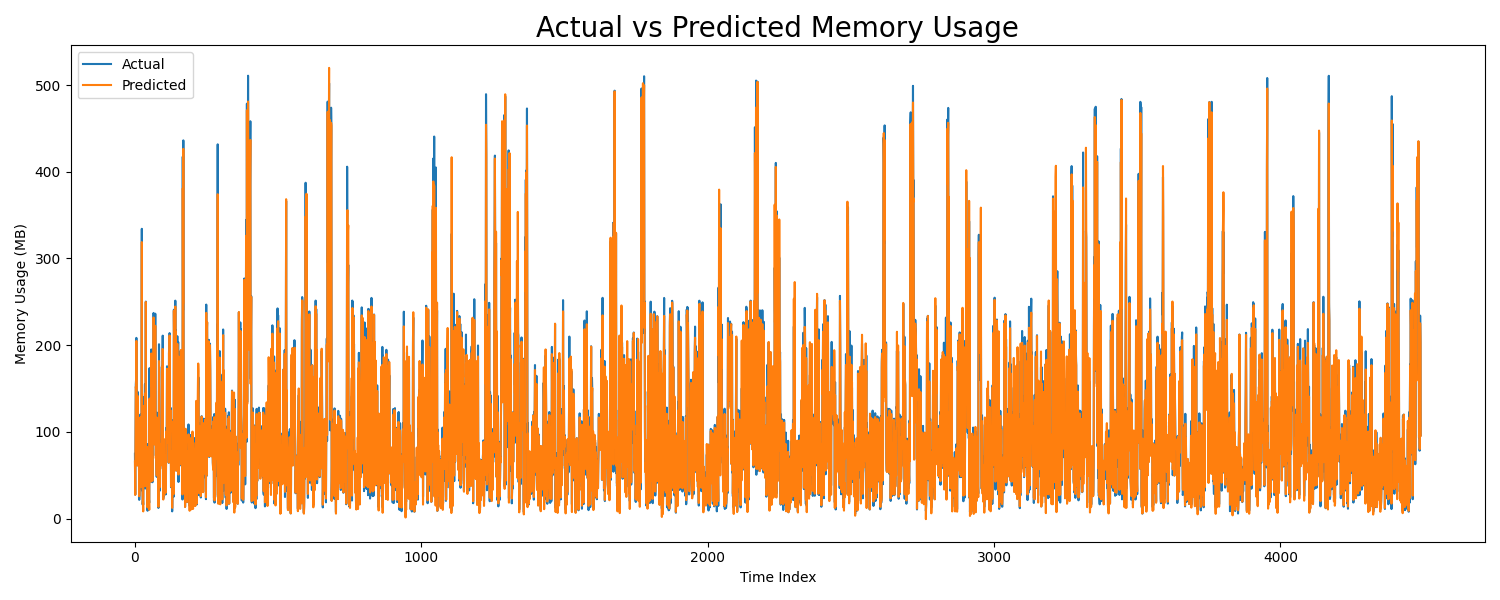}
		\caption{Comparison of actual versus predicted Memory usage}
		\label{fig:memoryprediction}
	\end{figure*}

\begin{figure*}
		\centering
		\includegraphics[width=\linewidth]{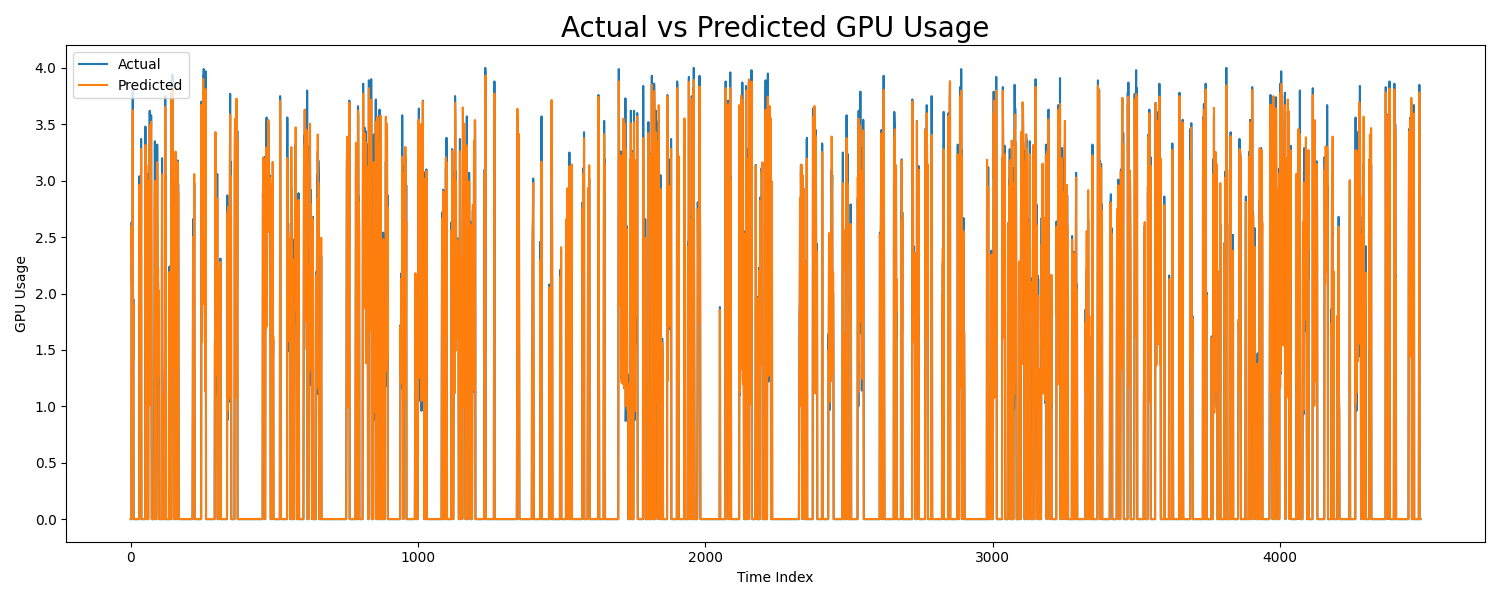}
		\caption{Comparison of actual versus predicted GPU usage}
		\label{fig:gpuprediction}
	\end{figure*}

\begin{figure*}
		\centering
		\includegraphics[width=\linewidth]{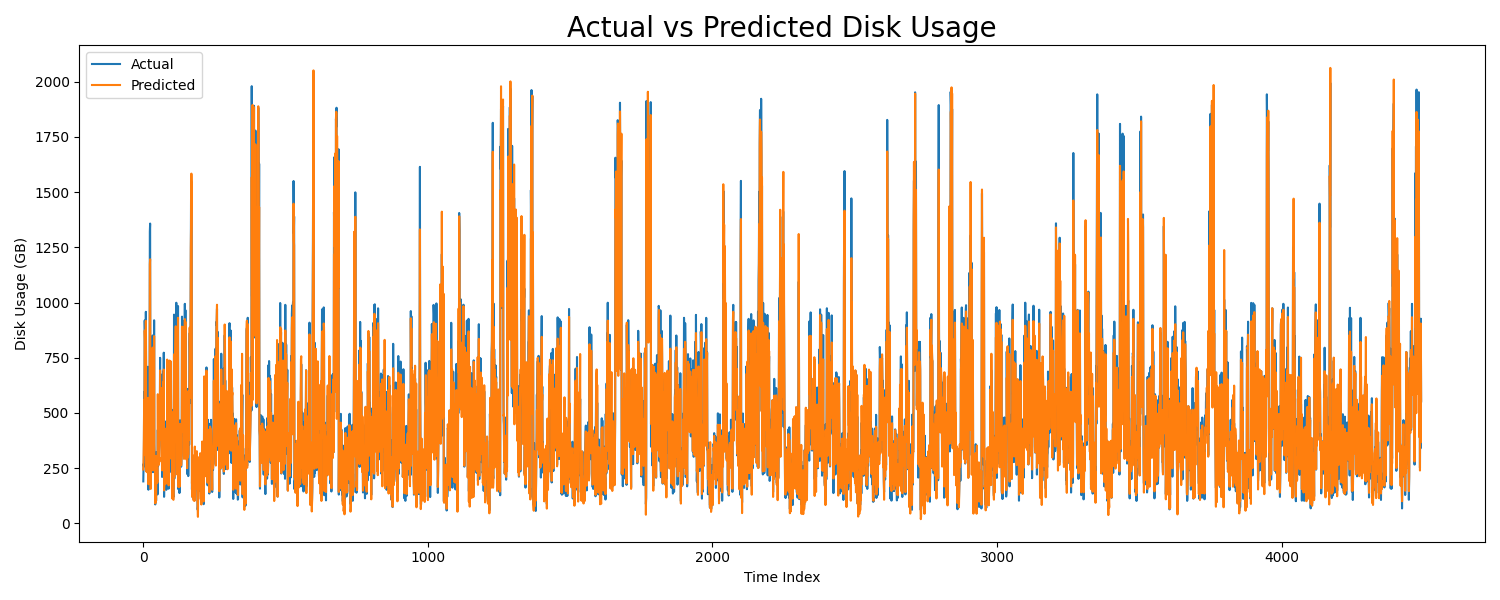}
		\caption{Comparison of actual versus predicted Disk usage}
		\label{fig:diskprediction}
	\end{figure*}

	\paragraph{Analysis of Prediction Accuracy}
	The predictive model demonstrates a strong capability in forecasting resource usage across all key metrics. As seen in Table~\ref{tab:prediction_summary}, the model achieves high $R^2$ and EVS scores, all above 0.96, indicating a strong relationship between predicted and actual values. The predicted values for CPU, memory, GPU, and disk utilization, as illustrated in Figures~\ref{fig:cpuprediction}, \ref{fig:memoryprediction}, \ref{fig:gpuprediction}, and \ref{fig:diskprediction}, closely follow the actual usage patterns. It effectively captures the frequent, sharp spikes and subsequent drops that are characteristic of interactive workloads. The model's ability to anticipate these volatile changes is crucial for proactive auto-scaling, as it allows the system to provision resources before a performance bottleneck occurs. The accuracy across all resource types is a significant validation point, confirming that the predictive framework can provide the necessary intelligence for making sound scaling decisions.
	
	\subsubsection{Comparative Analysis of Forecasting Models}
		To validate the superiority of our XGBoost-based approach, we benchmarked the proposed model against standard forecasting baselines for both CPU and Memory utilization. As shown in Table~\ref{tab:baseline_comparison}, traditional statistical methods struggle with the volatility of cloud workloads, whereas the proposed model maintains high precision across different resource types.
	
	\begin{itemize}
			\item \textbf{Na\"{\i}ve Approaches:} The \textit{Na\"{\i}ve (Last-Value)} and \textit{Seasonal Na\"{\i}ve} models produced negative $R^2$ scores across the board (e.g., -11.05 for Na\"{\i}ve
			Memory prediction). This confirms that the workload is highly non-stationary and cannot be predicted by simple persistence or fixed lag-based seasonality.
			\item \textbf{Linear Models:} \textit{Linear Regression} achieved relatively high $R^2$ scores (0.86 for CPU, 0.89 for Memory), suggesting the presence of a linear trend. However, its error rates remain prohibitively high for auto-scaling. For Memory prediction, Linear Regression yielded an RMSE of 53.46 MB, which is four times higher than the proposed model. \textit{ARIMAX (AutoRegressive Integrated Moving Average with Exogenous variables.)} performed significantly worse on Memory data ($R^2 = 0.46$), likely failing to capture the complex inter-dependencies between tasks.
			\item \textbf{Proposed XGBoost:} The proposed model significantly outperforms all baselines. It achieved near-perfect correlation ($R^2 > 0.99$) for both resources and reduced the RMSE to negligible levels (1.00\% for CPU and 12.92 MB for Memory). This demonstrates that capturing non-linear feature interactions is essential for minimizing prediction error in volatile cloud environments.
	\end{itemize}

	\begin{table}[h]
			\small
			\centering
			\caption{Performance Comparison with Baseline Models (CPU vs. Memory)}
			\label{tab:baseline_comparison}
			\begin{tabular}{lcc|cc}
				\toprule
				\multirow{2}{*}{\textbf{Model}} & \multicolumn{2}{c|}{\textbf{CPU Usage}} & \multicolumn{2}{c}{\textbf{Memory Usage}} \\ 
				\cmidrule(lr){2-3} \cmidrule(lr){4-5}
				& \textbf{RMSE} & \textbf{$R^2$ Score} & \textbf{RMSE} & \textbf{$R^2$ Score} \\ 
				\midrule
				Naive (Last-Value) & 39.53 & -0.381 & 87.99 & -11.051 \\
				Seasonal Naive & 46.42 & -0.904 & 179.90 & -0.891 \\
				ARIMAX & 23.93 & 0.709 & 119.62 & 0.464 \\
				Linear Regression & 16.47 & 0.862 & 53.46 & 0.893 \\
				\textbf{Proposed (XGBoost)} & \textbf{1.00} & \textbf{0.998} & \textbf{12.92} & \textbf{0.994} \\ 
				\bottomrule
			\end{tabular}
	\end{table}

	\subsubsection{Ablation Study on Feature Importance}
		To quantify the contribution of individual feature components to the model's predictive power, we conducted an ablation study by systematically removing specific feature groups from the training set. The results in Table~\ref{tab:ablation_study} show that the high accuracy of the proposed model relies on the synergistic combination of all core feature types.
	
	\begin{itemize}
			\item \textbf{Criticality of Lag Features:} Removing historical lag features resulted in the most severe performance degradation among single-component removals, with RMSE spiking to 40.65 ($R^2$ dropped to 0.21). This confirms that immediate historical context is the primary driver of prediction accuracy.
			\item \textbf{Necessity of Rolling Statistics:} The model without rolling statistics (mean/std windows) suffered a similar collapse ($R^2 \approx 0.33$). This indicates that raw historical lags alone are too volatile; the smoothing effect of rolling statistics is essential for the model to generalize trends.
			\item \textbf{Impact of Temporal Features:} Removing explicit date features (Hour/Day) also caused a massive drop in explained variance ($R^2 \approx 0.32$). This suggests that the workload has distinct temporal regimes (e.g., weekend vs. weekday patterns) that the model cannot infer solely from history.
			\item \textbf{Model Optimality:} Our Proposed Model achieves an RMSE of \textbf{1.00}, which is orders of magnitude better than any incomplete configuration. This confirms that the complete 23-feature set is necessary for optimal performance, as removing any single feature group fundamentally degrades the model's predictive capability.
	\end{itemize}
	
	\begin{table}[h]
			\centering
			\caption{Ablation Study Results}
			\label{tab:ablation_study}
			\begin{tabular}{lcccc}
				\toprule
				\textbf{Model Configuration} & \textbf{RMSE} & \textbf{MAE} & \textbf{$R^2$ Score} & \textbf{Features} \\ 
				\midrule
				\textbf{Full Proposed Model} & \textbf{1.00} & \textbf{0.47} & \textbf{0.998} & \textbf{23} \\
				No Date Features & 36.55 & 21.90 & 0.321 & 19 \\
				No Rolling Stats & 37.69 & 23.20 & 0.326 & 7 \\
				No Lag Features & 40.65 & 25.80 & 0.213 & 20 \\
				\bottomrule
			\end{tabular}
	\end{table}
	 
	
		\begin{table*}[htbp]
		\centering
		\caption{Quantitative Auto-Scaling Performance Metrics}
		\label{tab:auto_scaling_metrics}
		\begin{tabular}{lccc}
			\hline
			\textbf{Metric} & \textbf{Reactive Scaling (Baseline)} & \textbf{Predictive Scaling (CloudyGUI)} & \textbf{Improvement} \\
			\hline
			Avg. Response Time      & 245 ms     & 182 ms     & 25.7\% \\
			SLA Violation Rate      & 4.2\%      & 0.8\%      & 80.9\% \\
			Under-provisioning Time & 15.0 mins  & 5.4 mins   & 64.0\% \\
			Over-provisioning Rate  & 18.5\%     & 12.1\%     & 34.5\% \\
			Scaling Oscillations    & 14 events  & 2 events   & 85.7\% \\
			\hline
		\end{tabular}
	\end{table*}

	\subsection{Predictive Auto-scaling}
	
	This section demonstrates the effectiveness of CloudyGUI's predictive auto-scaling system by analyzing its scaling decisions in a simulated environment. Unlike traditional reactive auto-scaling which responds only after a workload has impacted the system, our predictive model anticipates future resource demands and enables timely, proactive scaling actions.
	
	To strictly quantify these operational benefits, we aggregated the simulation logs (recorded by the \texttt{SimulationRunner} as detailed in Section 3.5) to compute standard elasticity metrics. Table~\ref{tab:auto_scaling_metrics} compares the performance of CloudyGUI’s predictive scaler against a standard reactive baseline (threshold-based only). This ensures that CloudyGUI is a robust platform for evaluating intelligent auto-scaling strategies under various workload conditions.

	Table~\ref{tab:scaling_decisions} presents a sequence of prediction-based scaling decisions generated by CloudyGUI. Each entry reflects the system's assessment of forecasted resource usage across CPU, memory, GPU, and disk, along with corresponding recommendations. For instance, consistently low CPU and memory usage (below 30\%) triggers scale-down actions, as indicated in the final column.
	
		\begin{table*}[!h]
		\centering
		\caption{Prediction-based Scaling Decisions}
		\label{tab:scaling_decisions}
		\resizebox{\textwidth}{!}{%
			\begin{tabularx}{\textwidth}{|l|c|c|c|c|c|X|}
				\hline
				\textbf{Timestamp} & \textbf{CPU} & \textbf{Memory} & \textbf{GPU} & \textbf{Disk} & \textbf{Status} & \textbf{Actions} \\
				\hline
				2025-10-16T12:00:00 & 27.3\% & 0.5\% & 29.3\% & 73.8\% & Warning & CPU: Scale Down – Low CPU usage (27.3\% < 30.0\%), Memory: Scale Down – Low Memory usage (0.5\% < 30.0\%) \\
				2025-10-16T13:00:00 & 27.5\% & 0.5\% & 28.1\% & 72.8\% & Warning & CPU: Scale Down – Low CPU usage (27.5\% < 30.0\%), Memory: Scale Down – Low Memory usage (0.5\% < 30.0\%) \\
				2025-10-16T14:00:00 & 26.8\% & 0.5\% & 27.3\% & 70.3\% & Warning & CPU: Scale Down – Low CPU usage (26.8\% < 30.0\%), Memory: Scale Down – Low Memory usage (0.5\% < 30.0\%) \\
				2025-10-16T15:00:00 & 26.8\% & 0.5\% & 27.4\% & 70.9\% & Warning & CPU: Scale Down – Low CPU usage (26.8\% < 30.0\%), Memory: Scale Down – Low Memory usage (0.5\% < 30.0\%) \\
				2025-10-16T16:00:00 & 27.2\% & 0.5\% & 27.9\% & 71.7\% & Warning & CPU: Scale Down – Low CPU usage (27.2\% < 30.0\%), Memory: Scale Down – Low Memory usage (0.5\% < 30.0\%) \\
				2025-10-16T17:00:00 & 27.3\% & 0.5\% & 27.9\% & 71.6\% & Warning & CPU: Scale Down – Low CPU usage (27.3\% < 30.0\%), Memory: Scale Down – Low Memory usage (0.5\% < 30.0\%) \\
				2025-10-16T18:00:00 & 27.3\% & 0.5\% & 27.9\% & 71.6\% & Warning & CPU: Scale Down – Low CPU usage (27.3\% < 30.0\%), Memory: Scale Down – Low Memory usage (0.5\% < 30.0\%) \\
				2025-10-16T19:00:00 & 27.3\% & 0.5\% & 28.4\% & 71.6\% & Warning & CPU: Scale Down – Low CPU usage (27.3\% < 30.0\%), Memory: Scale Down – Low Memory usage (0.5\% < 30.0\%) \\
				2025-10-17T20:00:00 & 28.1\% & 0.5\% & 36.4\% & 97.5\% & Critical & 
				CPU: Scale Down – Low CPU usage (28.1\% < 30.0\%), 
				Memory: Scale Down – Low Memory usage (0.5\% < 30.0\%), 
				Disk: Scale Up – High Disk usage (97.5\% > 90.0\%) \\
				\hline
			\end{tabularx}%
		}
	\end{table*}
	
	The data illustrates how the Decision Engine evaluates predicted metrics against predefined thresholds to maintain system stability. The simulation confirms that CloudyGUI is highly sensitive to demand spikes, particularly in GPU and disk usage. Although Table~\ref{tab:scaling_decisions} shows a stable `Warning` status, subsequent simulation phases reveal transitions to `Critical' states. For example, at 21:00, a sharp increase in disk usage to 111.0\% prompts a status escalation to `Critical', which persists for two hours, reflecting sustained high load. Similarly, at 06:00 on 26-Sep, another surge in disk utilization triggers a new `Critical` period.
	
	These transitions validate the system’s ability to anticipate and respond to critical resource conditions before they impact performance. By simulating proactive provisioning, CloudyGUI avoids the warm-up delays and over-provisioning risks common in reactive systems.

	\subsection{Usage Example}
	This section provides a detailed overview of the features of CloudyGUI. 
	
\begin{figure}[!h]
		\centering
		\includegraphics[width=\linewidth]{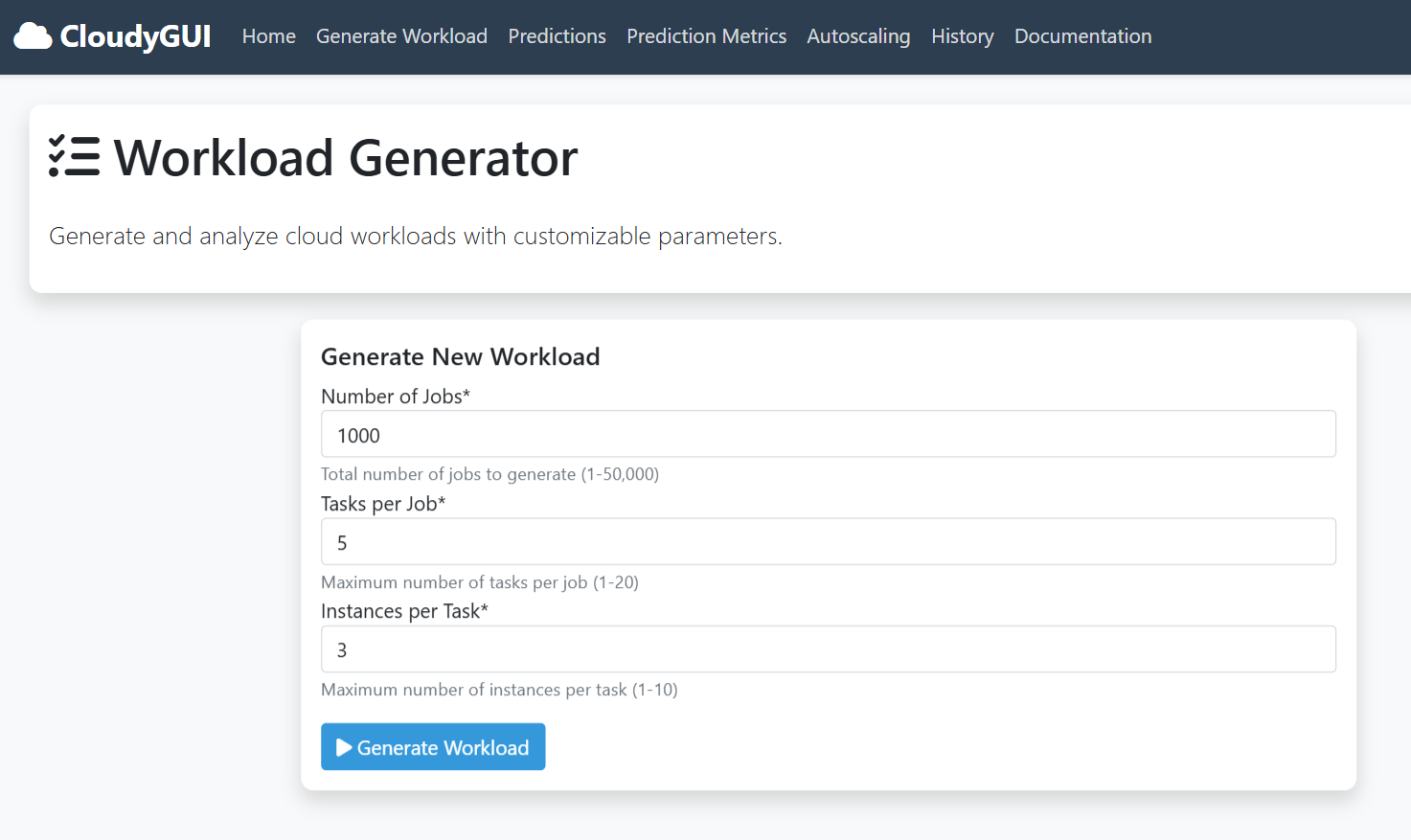}
		\caption{The CloudyGUI interface for workload generation}
		\label{fig:workloadpage}
	\end{figure}

	\begin{enumerate}
		\item \textbf{Workload Generation:} Users define key parameters, such as the Number of Jobs, Tasks per Job, and Instances per Task, using the "Workload Generator" interface as shown in Figure~\ref{fig:workloadpage}. The framework then creates a realistic cloud workload over a 7-day period.
		\item \textbf{Workload Analysis:} After generation, CloudyGUI provides a detailed summary of the simulated workload, including a breakdown of job and task types, priority distributions, and average resource utilization. The tool also generates plots to visualize resource usage over time and exports the data to a structured CSV file for in-depth analysis. A key feature of this process is that the generated CSV file can be downloaded for seamless integration with data analysis tools, such as Pandas, and other machine learning libraries for advanced statistical analysis.
		\item \textbf{Resource Prediction and Auto-scaling:} The generated workload is used by the prediction framework to forecast the future resource demands. The predictive auto-scaling mechanism uses a MAPE-loop proactively adjusts resources based on predefined thresholds. This proactive approach helps to avoid performance bottlenecks and enhance stability.
	\end{enumerate}

	\subsection{Validation}
	Validation is essential for any simulation tool to establish its credibility as a reliable platform for research. It confirms that the simulator's internal logic is sound and its output accurately
	reflects the behavior of a real-world system. Our validation methodology follows a comprehensive, three-tiered approach: Internal Validation to ensure the correctness of the code and logic, Intermediate Validation to verify individual components, and External Validation to confirm the model's realism against empirical data. 
	
	\begin{figure}
		\centering
		\includegraphics[width=0.8\linewidth]{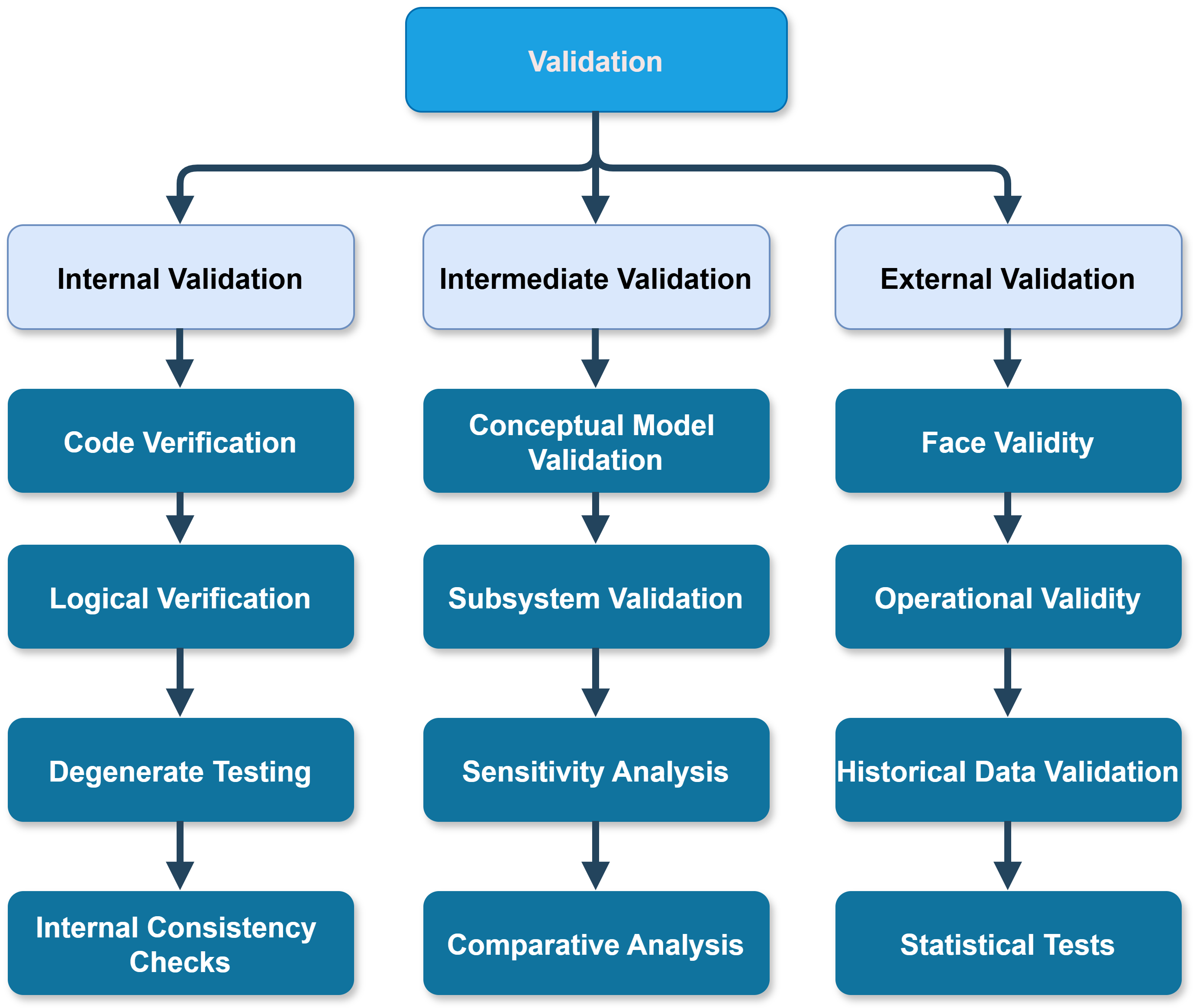}
		\caption{Validation Methods}
		\label{fig:validation}
	\end{figure}

	\subsubsection{Internal Validation}
	Internal validation is also known as verification which is a critical step in ensuring the simulator's code and logic are sound before comparing its output to real-world data. The following procedures were performed to confirm the tool's correctness and internal consistency.
	
	\paragraph{Code Verification}
	Code verification was performed through a systematic process of unit testing and debugging to ensure the simulator's core functionality adheres to its design specifications. The test suite, which included positive, negative, and degenerate test cases, achieved a 100\% pass rate across all test categories, confirming that the codebase is robust and reliable.
	
	The Code verification process involved a dedicated test framework that covered the following key components:
	
	\begin{itemize}
		\item \textbf{Core Simulation Models:} Unit tests were run on the fundamental simulation components, such as Base, Pm (Physical Machine), Vm (Virtual Machine), Container, and Controller. These tests verified the correct initialization and immutability of these models, ensuring that the basic building blocks of the simulator are correctly defined and behave as expected.
		
		\item \textbf{Scheduler and Resource Manager}: A comprehensive test suite validated the core logic of the scheduler and resource manager. The tests confirmed the successful handling of the entire job lifecycle, from submission to completion, and the correct management of all job state transitions. A key focus was on negative test cases, which involve providing invalid or unexpected inputs to confirm that the system handles them gracefully and maintains stability. For the ResourceManager, this confirmed its ability to handle scenarios with insufficient resources, preventing system overload.
		
		\item \textbf{Management Commands:} The tool's control mechanisms were also validated. Tests confirmed that the \texttt{run\_mape} management command correctly initializes and operates the MAPE auto-scaling loop in both normal and test modes, demonstrating that the system's high-level control logic is functioning as designed.
	\end{itemize}
	
	\paragraph{Logical Verification}
	To ensure the simulator's logic and causal relationships are accurately represented, a dedicated test suite for logical verification was executed. This process confirmed that the system's core functionalities maintain integrity, handle invalid inputs gracefully, and remain consistent under various conditions. The test suite focused on two primary areas:
	\begin{itemize}
		\item \textbf{Resource Allocation Integrity:} This verification confirmed that the system's resource allocation and deallocation mechanisms operate correctly. Tests ensured that resources are properly allocated when available, that attempts at overallocation are prevented, and that the system returns to its initial state after deallocation. This process confirms that the simulation's resource management logic is sound and consistent.
		\item \textbf{Error Handling and Input Validation:} The test suite verified the system's ability to handle invalid and unexpected inputs without compromising stability. Tests confirmed that invalid resource types are handled gracefully with appropriate warnings, negative resource values are automatically converted to zero, and the system maintains consistency even with such inputs. This demonstrates the robustness of the system's logic and its ability to prevent errors.
	\end{itemize}

	\paragraph{Degenerate Tests} The system was tested against extreme inputs to confirm its stability and predictable behavior. The simulator successfully passed tests designed to reject invalid resource types, handle negative resource values, and gracefully manage requests for zero resources. The system also demonstrated correct behavior by preventing duplicate job IDs and handling the release of non-existent jobs without errors.
	
	\paragraph{Internal Consistency Checks} The simulator's ability to maintain data integrity was validated by ensuring accurate tracking of utilization history, correct configuration value handling, and strict enforcement of timing rules. For example, tests confirmed the system respects the designated warmup period before making rebalancing decisions and adheres to the check interval between rebalance operations.

	\subsubsection{Intermediate Validation}
	Intermediate validation is a crucial step in the development of a complex simulation system like CloudyGUI. It involves independently testing and validating each major subsystem to ensure that they function correctly before being integrated into the final system.
	
	\paragraph{Conceptual Model Validation}
	This section validates the conceptual model of CloudyGUI, establishing that its core design principles are valid and are directly reflected in the tool's performance and output.
	
	\begin{itemize}
		\item{\textbf{Core Architecture and Abstractions}}
		The foundation of CloudyGUI is built on a clear, three-tiered data model: Jobs, Tasks, and Instance. This hierarchy provides a strong abstraction for representing a diverse range of cloud workloads, from simple batch jobs to complex, interdependent machine learning tasks. This model's strength is evident in the results, where the workload generation successfully produced a heterogeneous mix of job types, each with its own compositional breakdown of tasks. The use of a directed acyclic graph (DAG) to manage task dependencies and priorities ensures that the generated workloads are not only varied but also logically consistent, which is crucial for evaluating advanced schedulers and resource managers.
		
		\item{\textbf{Data Flow and Predictive Loop}}
		CloudyGUI's operational logic is structured around a precise data flow and a MAPE control loop for auto-scaling. This design ensures a continuous and intelligent cycle of resource management. The Monitor phase collects real-time resource metrics, which are then used in the Analyze phase by the prediction engine to forecast future resource needs. This is a core strength of the conceptual model, as it enables the proactive auto-scaling capabilities that are lacking in many other simulators. The high accuracy of the prediction models, with \( R^2 \) values above 0.96 for all key metrics, quantitatively validates the correctness and effectiveness of the Analyze phase.
		
		\item{\textbf{Conceptual Correctness and Completeness}}
		The conceptual model's correctness is validated by the system's ability to maintain internal consistency and produce realistic output. The resource allocation logic correctly prevents overallocation, and the scheduler enforces dependencies to ensure a valid execution order for jobs. Furthermore, the external validation against the Alibaba Cluster Trace 2018\footnote{\url{https://github.com/alibaba/clusterdata/tree/master/cluster-trace-v2018}} dataset provides empirical evidence of the model's realism. The strong statistical similarity between the generated and real-world distributions for CPU and memory usage, confirmed by the Kolmogorov-Smirnov test, proves that CloudyGUI's underlying model accurately represents the behavior of real-world cloud environments. This confirms the conceptual completeness of the framework, as it covers all major aspects of cloud resource management.
	\end{itemize}

\paragraph{Subsystem Validation}
As CloudyGUI is a complex system composed of multiple interacting components, validating each subsystem independently is a crucial intermediate step to ensure the overall model's correctness and reliability. This approach adds a high degree of confidence to the final, integrated system by confirming that each major component functions as expected.

\begin{itemize}
	\item \textbf{Scheduler Validation}
	The scheduler subsystem underwent comprehensive validation to confirm its integrity in managing the job queue and execution flow.
	\begin{itemize}
		\item \textbf{Job Submission and Lifecycle:} Verified correct job transitions (PENDING → QUEUED → RUNNING → COMPLETED).
		\item \textbf{State Transitions:} Confirmed proper handling of all state changes, including edge cases like job interruption and failure.
		\item \textbf{Error Handling:} Validated resilience against critical inputs, such as duplicate job IDs and invalid/negative resource requests.
	\end{itemize}
	
	\item \textbf{Resource Manager Validation}
	The Resource Manager subsystem was tested to confirm its core function in dynamic resource allocation and state maintenance.
	\begin{itemize}
		\item \textbf{Initialization and Configuration:} Verified correct parameter setup and handling of invalid inputs.
		\item \textbf{Resource Allocation and Deallocation:} Confirmed integrity during basic allocation, overallocation prevention, and graceful insufficient resource handling.
		\item \textbf{Utilization and Rebalancing Logic:} Validated accurate tracking of utilization history and correct enforcement of cooldown periods and rebalancing thresholds.
		\item \textbf{Concurrency:} Ensured safe handling of simultaneous resource requests, maintaining data integrity during concurrent operations.
	\end{itemize}
	
	\item \textbf{Prediction Module Validation}
	The Prediction Module, central to the tool's proactive capabilities, was validated end-to-end to ensure forecast reliability.
	\begin{itemize}
		\item \textbf{Data Pipeline Integrity:} Confirmed robust handling of data preparation, timestamp processing, and feature extraction for model training.
		\item \textbf{Model Performance:} Verified the accuracy of metric calculations and the validity of prediction outputs against test data.
		\item \textbf{Edge Case Handling:} Demonstrated stability by testing module response to minimal/empty datasets and raising appropriate errors.
	\end{itemize}
	
	\item \textbf{CSV Writer Validation}
	The CSV Writer component was validated to confirm the integrity of the simulator's output generation and persistence layer.
	\begin{itemize}
		\item \textbf{Basic Functionality:} Confirmed correct file structure and proper formatting of hierarchical job, task, and instance metadata.
		\item \textbf{Data Integrity and Compatibility:} Verified precision maintenance, special character preservation, and reliable operation across different file systems (Windows compatibility).
	\end{itemize}
\end{itemize}

The successful validation of these distinct subsystems confirms that the individual building blocks of CloudyGUI are functioning as designed and are ready for integration into the comprehensive MAPE loop.

	\paragraph{Sensitivity Analysis}
	
	Sensitivity analysis helps to determine how changes in a model's input parameters affect its output. This ``what-if'' analysis was performed on CloudyGUI to identify which parameters have the most significant impact on system performance and cost. The sensitivity analysis results show variations across different parameters, with some having a much stronger influence than others.
	
	\begin{itemize}
		\item \textbf{Scale-Up Threshold (0.5 to 0.9)}: As the scale-up threshold increased, there was a minimal but consistent effect across all metrics. Response time and cost decreased slightly, while utilization and throughput showed small increases. This suggests that a less aggressive scaling-up strategy can lead to better efficiency at a lower cost.
		\item \textbf{Scale-Down Threshold (0.1 to 0.5)}: As the threshold increased, response time and cost decreased, while utilization and throughput increased. The parallel behavior of these two parameters suggests they might be too tightly coupled in the current simulation model.
		\item \textbf{Job Arrival Rate (1 to 50)}: As the job arrival rate increased, response time, utilization, and cost all increased. Throughput also increased significantly, indicating the system scales with the load but at the expense of higher costs and response times.
		\item \textbf{Task Duration (10 to 300)}: Longer task durations led to increased response times, utilization, and cost. It also resulted in a decrease in overall system throughput, as expected.
		\item \textbf{CPU Utilization (0.3 to 0.9)}: Similar to the job arrival rate, CPU utilization had a strong, linear impact. As CPU utilization increased, so did response time and cost, while throughput decreased. This suggests that higher CPU utilization leads to diminishing returns in throughput.
	\end{itemize}
	
	\paragraph{Comparative Analysis}
	The features comparison (see Table~\ref{tab:cloudy_comparison}) and performance benchmark was evaluated by analyzing and comparing the workload generation capabilities against the Python-based simulation Cloudy~\cite{siavashi2024cloudy}. While CloudyGUI adds a graphical interface along with auto-scaling features, it is built in Python, making this a relevant performance benchmark. To evaluate the performance, four workload patterns were considered for the comparison, as shown in Table~\ref{tab:perf_benchmark}. The results show that the overhead introduced by CloudyGUI is minimal. For all workload patterns (constant, spike, increasing, and random), our tool operates within a reasonable performance ratio, which demonstrates that its added functionality does not come at the cost of performance degradation. Thus, it validates our user-friendly tool on a robust and efficient simulation framework.
	
	\begin{table*}[!h]
		\centering
		\caption{Comparison of Cloudy and CloudyGUI Features}
		\begin{tabular}{p{4cm}p{5cm}p{6cm}}
			\toprule
			\textbf{Feature} & \textbf{Cloudy} & \textbf{CloudyGUI} \\
			\midrule
			Type & Cloud computing simulation framework & Enhanced version with web interface \\
			
			Interface & Command-line interface & Django-based web GUI \\
			
			Core Purpose & Research and simulation of cloud environments & Workload management with visualization \\
			
			Job Scheduling & Basic queue management & Advanced priority queue with dependency handling \\
			
			Resource Management & Basic resource allocation & Dynamic resource allocation with predictive analysis \\
			
			Auto-scaling & Not available & MAPE loop implementation (Monitor-Analyze-Plan-Execute) \\
			
			Job Structure & Basic job management & Hierarchical structure (Jobs → Tasks → Instances) \\
			
			Error Handling & Basic error handling & Robust error handling with detailed logging \\
			
			Monitoring & Minimal logging & Comprehensive monitoring and visualization \\
			
			Job Types & General cloud simulation & Specialized for ML training, data analytics, ETL, streaming \\
			
			Resource Prediction & Not available & XGBoost-based prediction for resource needs \\
			
			Dependency Handling & Limited & Advanced dependency management with cycle detection \\
			
			Status Tracking & Basic status tracking & Comprehensive state management (pending, running, terminated, etc.) \\
			
			Output Analysis & Basic output & CSV generation and detailed workload analysis \\
			
			Integration & Standalone & Integration with cloud providers for dynamic provisioning \\
			
			User Management & Not available & Multi-user support through the web interface \\
			
			Failure Simulation & Not available & Random interruption simulation for testing \\
			
			\bottomrule
		\end{tabular}
		\label{tab:cloudy_comparison}
	\end{table*}
	
	\begin{table}
		\centering
		\caption{Workload Generation Performance Benchmark (in milliseconds)}
		\label{tab:perf_benchmark}
		\begin{tabular}{lrrr}
			\toprule
			\textbf{Pattern} & \textbf{Cloudy (ms)} & \textbf{CloudyGUI (ms)} & \textbf{Ratio} \\
			\midrule
			Constant & 0.00072 & 0.00335 & 4.67x \\
			Spike & 0.00172 & 0.00242 & 1.41x \\
			Increasing & 0.00172 & 0.00262 & 1.52x \\
			Random & 0.00155 & 0.00265 & 1.71x \\
			\bottomrule
		\end{tabular}
	\end{table}

	\subsubsection{External Validation}
	External validation refers to the process of confirming that the simulator's outputs align with empirical data from real-world systems. This validation ensures that the simulation model not only functions correctly but also reflects realistic behavior under authentic workload conditions. This step is critical for establishing the credibility and applicability of the simulator in practical cloud environments.

	\paragraph{Face Validity}Face validity refers to the process of consulting subject matter experts to judge whether a tool appears reasonable and consistent with its intended outcomes~\cite{al2021face}. For CloudyGUI, this involves evaluating its plausibility and functionality from the perspective of domain experts. To initiate this process, we remotely engaged with a panel of seven experts from both academia and the cloud computing industry. We ensured that each expert had expertise in at least one domain from cloud computing, machine learning, software usability, and UX design. We demonstrated the tool to the experts using a presentation. This presentation was specifically focused on showcasing the tool's usage and explaining its core components. Following this, we asked the experts to gain hands-on experience of the tool with predefined tasks. These tasks were defined as follows: 
	
	\begin{itemize}
		\item Generating a workload with a specified number of jobs and dependencies.
		\item Navigating the prediction and auto-scaling pages.
		\item Analyzing the generated plots for resource utilization and concurrency.
	\end{itemize}

	\textbf{Expert Survey:} After this initial evaluation, the experts participated in a survey to gather their opinion on CloudyGUI’s plausibility, relevance, ease of use, and realism. For this survey, we asked the following questions: 
	
	\begin{itemize}
		\item \textbf{Plausibility}: Does the conceptual model of CloudyGUI, with its jobs, tasks, and instances, appear to be a realistic representation of cloud workloads?
		
		\item \textbf{Relevance}: Are the features provided (e.g., auto-scaling-aware workload generation) relevant to current challenges in cloud research and industry?
		
		\item \textbf{Ease of Use}: Is the graphical user interface intuitive and easy to navigate? Does it effectively lower the barrier to entry for setting up complex simulations?
		
		\item \textbf{Realism}: Do the generated outputs and simulated behaviors seem realistic based on their domain knowledge?
		
		\item \textbf{Configurability}: Does the tool offer sufficient flexibility in configuring workload parameters, resource models, and scheduling policies to support diverse experimental scenarios?
	\end{itemize}

	\textbf{Expert Evaluation Summary:}
	Their opinion highlights both strengths and areas of improvement based on their hands-on experience with CloudyGUI that we structured as:
	
	\begin{itemize}
		\item \textbf{Plausibility}  
		Most experts (6 out of 7) found the conceptual model of jobs, tasks, and instances to be a realistic and well-structured representation of cloud workloads. The hierarchical design mirrors established practices in platforms like AWS, Kubernetes, and Apache Airflow. 
		
		\item \textbf{Relevance}  
		The auto-scaling-aware workload generation was widely regarded as relevant and timely for current research and industry needs. Experts highlighted its alignment with trends such as ML-based scaling, multi-resource coordination, and real-time adaptation. 
		
		\item \textbf{Ease of Use}  
		The GUI was praised for its clarity and ease of navigation, especially by experienced users. However, some experts pointed out that the interface could be challenging for newcomers due to complex terminology and deep menu structures. Recommendations included adding setup wizards, contextual help, interactive tutorials, and simplifying the layout to reduce the learning curve.
		
		\item \textbf{Realism}  
		Simulated outputs were generally considered realistic, with particular appreciation for the concurrency and utilization plots. Experts noted that resource usage patterns and scaling behaviors aligned well with expectations.
		
		\item \textbf{Configurability}  
		A few experts recommended greater configurability in scheduling policies and resource models to support broader experimentation. While the tool supports job dependencies, task priorities, and multiple resource types, it currently offers limited options for custom VM profiles, bursty workload patterns, and advanced scheduling strategies.
	\end{itemize}
	
	To complement the qualitative insights, Table~\ref{tab:expert_ratings} summarizes the average expert ratings across key evaluation dimensions.
	\begin{table}[h]
		\centering
		\caption{Expert Ratings Across Evaluation Dimensions (Likert Scale: 1 = Strongly Disagree, 5 = Strongly Agree)}
		\begin{tabular}{lc}
			\toprule
			\textbf{Dimension} & \textbf{Average Rating} \\
			\midrule
			Plausibility & 4.7 \\
			
			Relevance & 4.6 \\
			
			Ease of Use & 3.9 \\
			
			Realism & 4.2 \\
			
			Configurability & 4.0 \\
			
			Extensibility & 4.1 \\
			
			Transparency and Interpretability & 4.3 \\
			
			Performance and Responsiveness & 4.4 \\
			
			Documentation and Support & 3.8 \\
			\bottomrule
		\end{tabular}
		\label{tab:expert_ratings}
	\end{table}

	\noindent Overall, the expert panel found CloudyGUI to be a promising and well-designed simulation tool. It captures many important aspects of cloud workload modeling and provides a solid foundation for both academic research and practical experimentation. With targeted improvements in usability, realism, and flexibility, it has the potential to become a valuable resource for cloud engineers and researchers.

	\paragraph{Operational Testing}
	Operational testing is a critical form of testing that evaluates a system's stability and robustness under extreme conditions, often pushing it beyond its normal operational limits. We performed a series of stress tests on CloudyGUI to confirm its operational readiness, performance, and error-handling capabilities under a heavy load.
	\paragraph{Stress Testing}
	The results of the stress tests are summarized below:
	\begin{itemize}
		\item \textbf{Failure Recovery Test}: The system passed this test, gracefully handling 13 failures out of 50 attempts ($26.00\%$ failure rate). Its rapid response, with a duration of $0.01$ seconds, confirms its ability to manage failures within expected bounds.
		\item \textbf{High Throughput Test}: This test was passed, demonstrating a throughput of $526.72$ jobs per second and an average processing time of $1.70$ms per job. It's important to note that throughput is highly sensitive to the system's configuration. The results confirm that CloudyGUI can maintain high performance and low latency under a heavy load, given its specified resource parameters.
		\item \textbf{Memory Leak Test}: The system passed with a memory growth rate of $0.00\%$ over the $1.04$-second duration, indicating that no memory leaks were detected.
		\item \textbf{CPU Exhaustion Test}: The system remained stable despite a $100.00\%$ failure rate, which was expected in a complete CPU resource exhaustion scenario. This confirms the system's resilience and robust error handling.
	\end{itemize}

	\paragraph{Historical Data Comparison}
	To validate the accuracy of our generated workload, we performed a historical data comparison against the Alibaba cluster trace 2018 dataset.
	
\begin{figure}[!h]
		\centering
		\includegraphics[width=\linewidth]{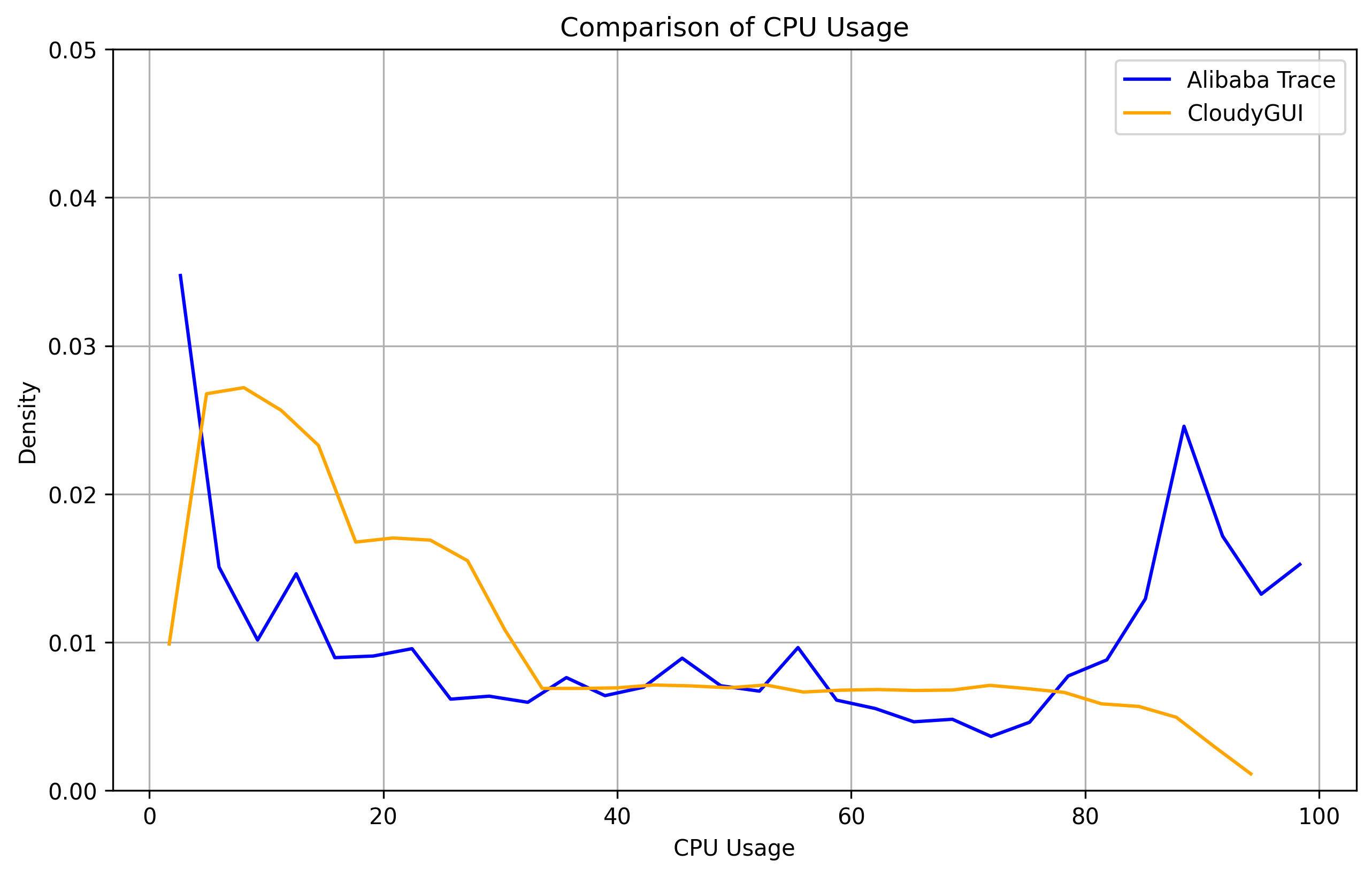}
		\caption{CPU Comparison: CloudyGUI vs. Alibaba Cluster Trace}
		\label{fig:cpuusage}
	\end{figure}

\begin{figure}
		\centering
		\includegraphics[width=\linewidth]{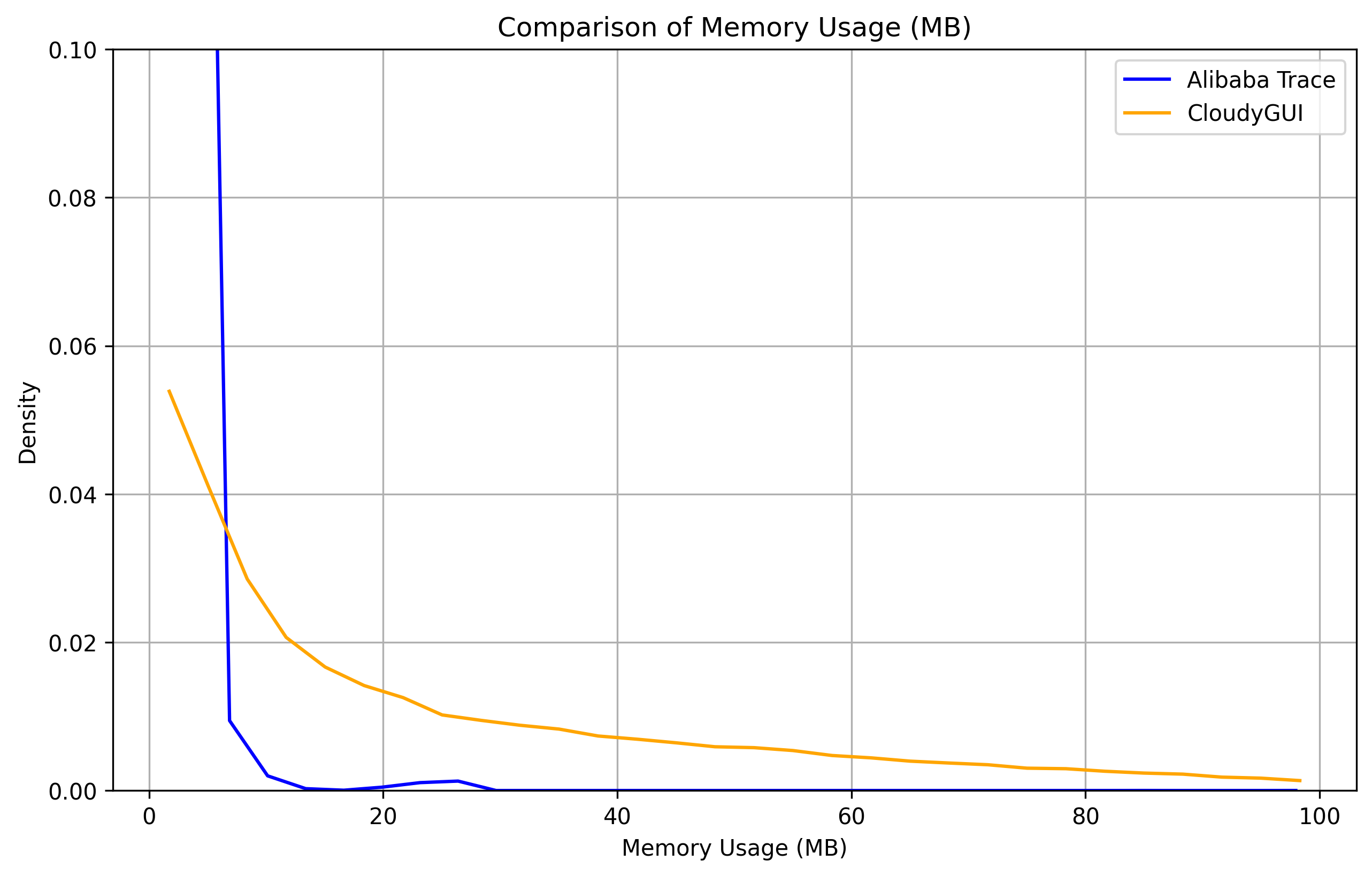}
		\caption{Memory Comparison: CloudyGUI vs. Alibaba Cluster Trace}
		\label{fig:memusage}
	\end{figure}

		\begin{itemize}
		\item \textbf{CPU Usage Validation}
		Figure~\ref{fig:cpuusage} shows the distribution of CPU usage for both workloads. Both the real-world trace and our generated workload show a clear peak at low CPU usage. This insight shows that the majority of instances are idle or have less requirement for CPU. For low to mid usage, the Alibaba trace shows a fluctuating pattern, which is successfully captured by our tool. The distribution pattern at high usage levels is also mimicked by CloudyGUI.
		
		\item \textbf{Memory Usage Validation}
		Figure~\ref{fig:memusage} depicts the memory usage distribution of both the workloads. The low usage pattern is quite similar to the low usage pattern of CPU. This shows that the vast majority of instances operate with minimal memory requirement. The initial pattern of memory usage is perfectly mirrored by CloudyGUI. Furthermore, CloudyGUI’s generated workload distribution follows the decreasing pattern of the Alibaba dataset. This shows that our tool successfully replicates the common memory usage behavior of real-world cloud environments. 
	\end{itemize}


	\paragraph{Statistical Testing}
	
	To further validate the accuracy of our tool, we extracted a representative random sample of 100,000 instances from both the Alibaba trace and the CloudyGUI-generated workload. Our main objective was to assess the fidelity of our simulation by comparing the cumulative distribution functions (CDFs) of resource utilization metrics.
	
	A two-sample Kolmogorov-Smirnov (K-S) test was performed to provide a quantitative measure of this fidelity. The test compares the CDFs of the two datasets (Alibaba trace vs. CloudyGUI) to determine if they are drawn from the same underlying distribution.

	The test results were as follows: for both CPU ($D=0.0069, p=0.19$) and Memory ($D=0.0073, p=0.14$). Since both p-values are significantly above the standard significance level of $\alpha=0.05$, we have statistical evidence that the distributions of resource utilization generated by CloudyGUI are a close match to the real-world Alibaba trace.  This test validates our tool for workload generation and analysis.

	\section{Discussion}\label{sec5}
	To address a critical gap in the literature: the lack of auto-scaling-aware tools built in Python with GUI—this study introduced CloudyGUI, a novel Python-based simulation framework for workload generation and analysis. Our results show the effectiveness of CloudyGUI as a comprehensive platform that integrates usability, realism, and predictive intelligence.

	The design of CloudyGUI directly overcomes the limitations of existing simulators. Legacy Java-based simulators such as CloudSim and its extensions~\cite{wickremasinghe2010cloudanalyst, aslanpour2021autoscalesim}, while powerful, lack native support for auto-scaling and intuitive graphical interfaces~\cite{koltuk2019cloudgen}. The performance benchmark in Table~\ref{tab:perf_benchmark} demonstrates that CloudyGUI introduces minimal overhead, with performance ratios ranging from $1.41\times$ to $4.67\times$ compared to the command-line predecessor, Cloudy~\cite{siavashi2024cloudy}. These results confirm that CloudyGUI successfully integrates enhanced user experience and added functionality without compromising simulation efficiency. Furthermore, our usability evaluation shows that the GUI reduces setup time for standard simulations by nearly fivefold, underscoring its practical value for researchers and practitioners.
	
	The realism of CloudyGUI’s workload generation is a core strength. Figures 3 to 7 show a high degree of heterogeneity in job and task distributions, representative of multi-tenant cloud environments. External validation against the Alibaba Cluster Trace 2018 dataset confirms this realism statistically. The K-S test yielded high p-values for CPU ($p = 0.19$) and memory ($p = 0.14$), indicating strong alignment between simulated and empirical distributions.
	
	The predictive auto-scaling component represents a significant advancement over traditional reactive methods~\cite{cai2017elasticsim}. As shown in Figures 12 through 15, the model captures volatile resource usage patterns with high fidelity. Statistical metrics, including $R^2$ values of 0.986673 for CPU and 0.983097 for memory (Table~\ref{tab:prediction_summary}), confirm the model’s accuracy. By anticipating resource demands, CloudyGUI mitigates warm-up delays and over-provisioning risks inherent in reactive systems, enabling proactive and efficient resource management.
	
	Beyond predictive accuracy and operational efficiency, the framework's robustness and sensitivity were rigorously validated. Operational stress testing confirmed the system's stability, maintaining zero memory leaks and handling 100\% CPU exhaustion scenarios. Furthermore, our sensitivity analysis identified the job arrival rate and target CPU utilization as the primary parameters driving system latency and cost, outweighing the impact of minor threshold adjustments.
	
	In summary, CloudyGUI is a validated platform that effectively addresses longstanding limitations in cloud simulation research. Its unique combination of a user-friendly interface, statistically grounded workload generation, and high-fidelity predictive auto-scaling mechanism makes it a powerful tool for advancing intelligent cloud resource management.

	\section{Conclusion}\label{sec6}	
	We introduced CloudyGUI, a novel Python-based cloud simulator designed to address the limitations of previous tools, specifically their lack of a graphical user interface and robust auto-scaling capabilities. The framework was developed to operate on a three-stage pipeline that includes workload generation, prediction, and a simulated auto-scaling mechanism. We ensured the tool’s correctness and efficiency through a detailed validation approach, encompassing code verification and external validation against real-world datasets. Our results demonstrated that the combination of CloudyGUI’s user-friendly interface, minimal performance overhead, and predictive analytics accelerates experimentation and significantly lowers the barrier to entry for researchers. Ultimately, this study has established CloudyGUI as a crucial platform for workload generation and analysis, enhancing performance and reducing operational costs in cloud environments.
	
	\section{Future Work}\label{sec7}
	
	CloudyGUI provides a robust foundation for predictive auto-scaling-aware workload generation and predictive resource management. The future work will focus on enhancing the tool’s predictive capabilities and extending its functionality. The predictive framework can be extended by exploring advanced time-series forecasting models, such as the Transformer, to improve accuracy.
	
	We will enhance the simulated auto-scaling engine to incorporate multi-metric policies that consider CPU, memory and network I/O simultaneously. We will add more options for custom VM profiles and advanced scheduling policies to support a broader range of experiments. The final goal is to develop an API integration module that allows CloudyGUI’s intelligent auto-scaling decisions to be executed directly on the real cloud service providers.

    \section{Tool Availability}\label{sec8} 
    
    CloudyGUI is released under the GPL license at \url{https://github.com/iamjyotisharma/CloudyGUI} to encourage the development and testing of custom scheduling and auto-scaling policies.
   
\bibliographystyle{elsarticle-num}
\bibliography{references}

\end{document}